\documentclass[twocolumn]{aastex62}

\usepackage{savesym}
\savesymbol{tablenum}
\usepackage{siunitx}
\restoresymbol{SIX}{tablenum}

\usepackage{amsmath,bm}
\newcommand{ \angstrom}{\textup{\AA}}

\graphicspath{{./}{figures/}}
\usepackage[T1]{fontenc}

\usepackage{amssymb}

\submitjournal{ApJ}

\shorttitle{A 40-billion solar mass black hole in the extreme core of Holm~15A}
\shortauthors{Mehrgan et al.}

\begin{document}

\title{A 40-BILLION SOLAR MASS BLACK HOLE IN THE EXTREME CORE OF HOLM~15A,\\ THE CENTRAL GALAXY OF ABELL 85}

\author{Kianusch Mehrgan}
\affil{Max-Planck-Institut f\"ur extraterrestrische Physik, Giessenbachstrasse, D-85748 Garching} \affil{Universit\"ats-Sternwarte M\"unchen, Scheinerstrasse 1, D-81679 M\"unchen, Germany}
\email{kmehrgan@mpe.mpg.de}
\author{Jens Thomas}
\affil{Max-Planck-Institut f\"ur extraterrestrische Physik, Giessenbachstrasse, D-85748 Garching} \affil{Universit\"ats-Sternwarte M\"unchen, Scheinerstrasse 1, D-81679 M\"unchen, Germany}

\author{Roberto Saglia}
\affil{Max-Planck-Institut f\"ur extraterrestrische Physik, Giessenbachstrasse, D-85748 Garching} \affil{Universit\"ats-Sternwarte M\"unchen, Scheinerstrasse 1, D-81679 M\"unchen, Germany}

\author{Ximena Mazzalay}
\affil{Max-Planck-Institut f\"ur extraterrestrische Physik, Giessenbachstrasse, D-85748 Garching} \affil{Universit\"ats-Sternwarte M\"unchen, Scheinerstrasse 1, D-81679 M\"unchen, Germany}

\author{Peter Erwin}
\affil{Max-Planck-Institut f\"ur extraterrestrische Physik, Giessenbachstrasse, D-85748 Garching} \affil{Universit\"ats-Sternwarte M\"unchen, Scheinerstrasse 1, D-81679 M\"unchen, Germany}

\author{Ralf Bender }
\affil{Max-Planck-Institut f\"ur extraterrestrische Physik, Giessenbachstrasse, D-85748 Garching} \affil{Universit\"ats-Sternwarte M\"unchen, Scheinerstrasse 1, D-81679 M\"unchen, Germany}

\author{Matthias Kluge}
\affil{Max-Planck-Institut f\"ur extraterrestrische Physik, Giessenbachstrasse, D-85748 Garching} \affil{Universit\"ats-Sternwarte M\"unchen, Scheinerstrasse 1, D-81679 M\"unchen, Germany}

\author{Maximilian Fabricius}
\affil{Max-Planck-Institut f\"ur extraterrestrische Physik, Giessenbachstrasse, D-85748 Garching} \affil{Universit\"ats-Sternwarte M\"unchen, Scheinerstrasse 1, D-81679 M\"unchen, Germany}

\begin{abstract}
Holm~15A, the brightest cluster galaxy (BCG) of the galaxy cluster Abell~85, has an ultra-diffuse
central region, $\sim\SI{2}{mag}$ fainter than the faintest depleted
core of any early-type galaxy (ETG) that has been dynamically
modelled in detail.  
We use orbit-based, axisymmetric Schwarzschild models to analyse the stellar
kinematics of Holm~15A from new high-resolution,
wide-field spectral observations obtained with MUSE at the VLT. We
find a supermassive black hole (SMBH) with a mass of
$(4.0 \pm 0.80) \times 10^{10} \ \si{M_{\odot}}$ at the center of Holm
15A. This is the most massive black hole with a direct dynamical
detection in the local universe.
We find that the distribution of
stellar orbits is increasingly biased towards tangential motions
inside the core. However, the tangential bias is less than in other cored elliptical
galaxies.
We compare Holm~15A with N-body simulations of mergers between galaxies with black holes and find that the observed amount of tangential
anisotropy and the shape of the light profile are consistent with a formation scenario where Holm~15A is the
remnant of a merger between two ETGs with pre-existing
depleted cores. 
We find that black hole masses in cored galaxies, including Holm~15A, scale inversely with the central stellar surface brightness and mass density, respectively. These correlations are independent of a specific parameterization of the light profile.

\end{abstract}

\keywords{galaxies: supermassive black holes -- galaxies: ETG and lenticular, cD 
-- galaxies: evolution -- galaxies: formation --stars: kinematics and dynamics
-- galaxies: center  -- clusters: individual (Abell~85)}

\section{Introduction} 
\label{sec:intro}
Holm~15A is the brightest cluster galaxy (BCG) of Abell~85. It is a very
luminous ($M_{V} = \SI{-24.8}{mag}$, \citealt{Kluge2019}) early-type
galaxy (ETG) with a high stellar mass of 
$M_{\star} \gtrsim 2 \times 10^{12} \ \si{M_{\odot}}$.  The rotational velocity of Holm~15A is
$v_{rot} \lesssim \SI{40}{km/s}$ and small compared to the velocity
dispersion $\sigma \sim \SI{350}{km/s}$. This is very common among
massive ETGs \citep[e.g][]{2011MNRAS.414..888E, 2016ARA&A..54..597C,
  2017MNRAS.464..356V}.  Despite its high overall luminosity, Holm~15A
has one of the faintest known central regions of any massive galaxy.

\begin{figure}
 \includegraphics[width=0.9\columnwidth]{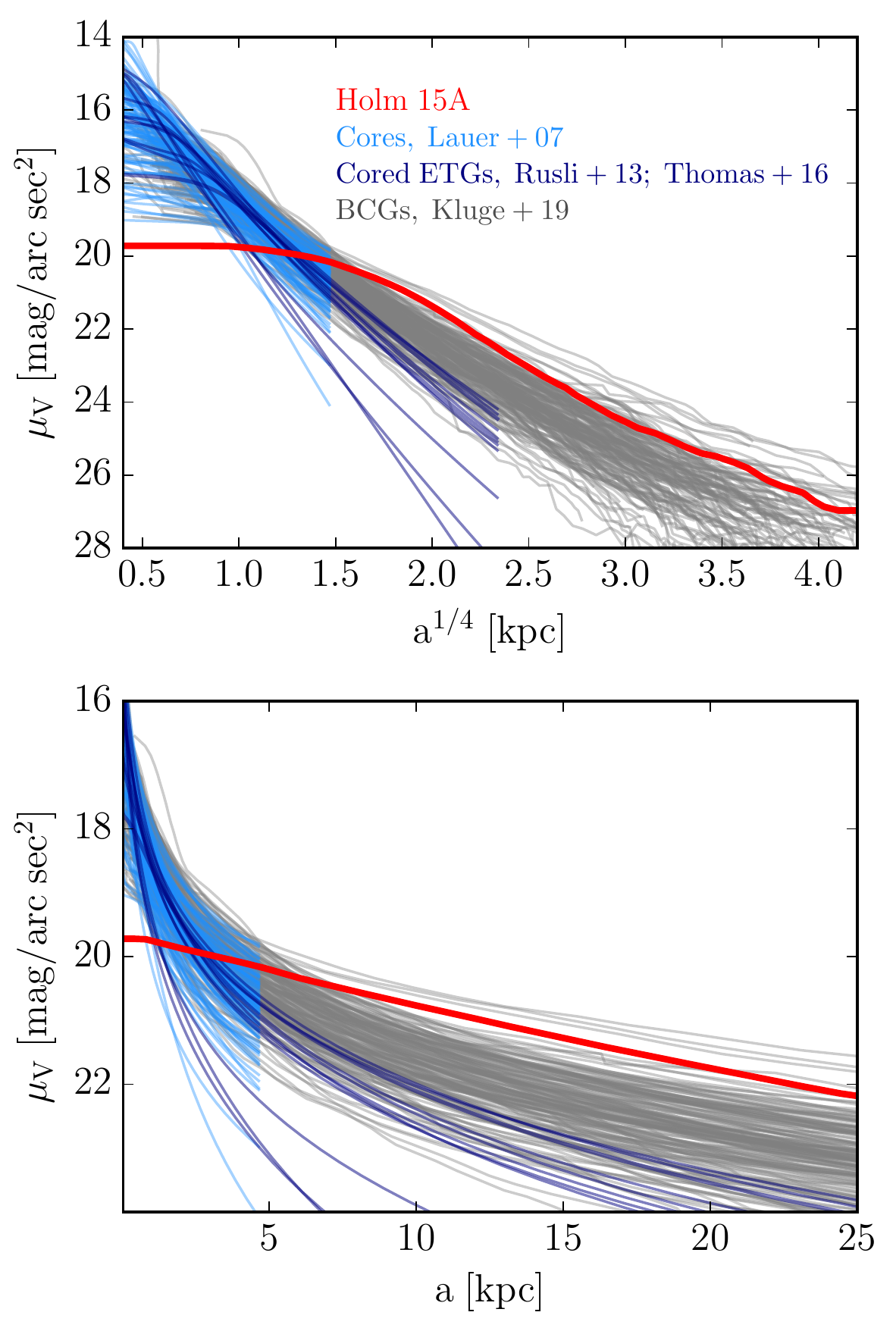}
 \caption{V-Band surface brightness profile of Holm~15A compared to
   the central $\SI{5}{kpc}$ of Nuker models of cored ETGs from
   \citet{2007ApJ...664..226L} (light blue), core-S\'{e}rsic models of
   cored ETGs with dynamical SMBH detections from
   \citet{2013AJ....146..160R} and \citet{2016Natur.532..340T} (dark
   blue), as well as observed light profiles of the 170 local BCGs
   from \citet{Kluge2019} (gray) over major axis.  Holm~15A's light profile has been
   shifted from $g'$-band assuming g-V = \SI{0.45}{mag} \citep{Kluge2019}, a K-correction of \SI{0.13}{mag}, cosmological dimming of \SI{0.23}{mag} and a galactic
   extinction of $A_{g} =0.125$mag.}
   \label{fig:holmvlauer}
\end{figure}

Figure \ref{fig:holmvlauer} compares Holm 15A's observed light profile
with Nuker models of the centers of cored ETGs from the
\citet{2007ApJ...664..226L} sample, core-S\'{e}rsic models of cored
ETGs with existing dynamical models from \citet{2013AJ....146..160R}
and \citet{2016Natur.532..340T}, as well as non-parametric light
profiles of BCGs from \citet{Kluge2019}. Evidently, at radii
$r \gtrsim \SI{30}{kpc}$ Holm15A's surface brightness profile is
characterised by a local S\'{e}rsic index $n \gtrsim 4$, typical for
massive ETGs and BCGs. Holm~15A is very bright though: only a handful of other BCGs have a higher surface brightness outside the central region ($r \gtrsim \SI{5}{kpc}$).

It is all the more striking then {\it how faint} the center of
Holm~15A is compared to ETGs from all three samples, BCG or
not. Indeed, among the 88 core galaxies in the
\citet{2007ApJ...664..226L} sample, the faintest center is still
$\sim 0.5 \ \si{mag/arcsec^2}$ brighter than the center of Holm~15A.
Among galaxies with detailed dynamical models, the difference is even
larger: $\sim 2 \ \si{mag/arcsec^2}$ (\citealp{2013AJ....146...45R},
\citealp{2016Natur.532..340T}, cf. Figure \ref{fig:holmvlauer}).

Such diffuse, shallow central surface brightness regions are commonly
referred to as `cores' and have been observed in massive early-type
galaxies (ETGs) for a long a time \citep[e.g.][]{1985ApJ...292..104L,
  1985ApJ...292L...9K, 1987nngp.proc..175F}. As methods for the
dynamical detection of supermassive black holes (SMBHs) of ETGs have
grown more sophisticated in recent years,
several tight scaling relations between
core properties and central black holes have been established.  In
particular, the most massive black holes in the local universe are
expected to be found in the centers of the largest, faintest cores
\citep[e.g.][]{1997AJ....114.1771F, 2007ApJ...664..226L,
  2013AJ....146..160R,2013ARA&A..51..511K,2016Natur.532..340T}.

The contemporary view of the formation of cores in massive ETGs is
that their observed properties are best explained via so-called
black hole binary `core scouring'.  Core scouring is driven by the
hardening of a SMBH binary naturally formed during dissipationless
mergers between ETGs which are thought to dominate the late growth
processes of massive galaxies
\citep[e.g.][]{2003ApJ...597L.117K,2006ApJ...636L..81N,
  boylan06,delucia06,2010ApJ...725.2312O}.  Gravitational slingshots
eject stars on predominantly radial orbits from the center of the
remnant galaxy, producing a cored central light profile
\citep[e.g.][]{1980Natur.287..307B, 1980AJ.....85.1281H,
  1991Natur.354..212E, 2004AJ....127.1917T,2001ApJ...563...34M,
  2003ApJ...582..559V, 2005LRR.....8....8M,
  merritt06,2013degn.book.....M, 2013AJ....146..160R,
  2018ApJ...864..113R}. This core-formation channel can explain the
fundamental characteristics of core galaxies: (1) the observed uniform
tangentially biased orbit structure in cores
\citep{2001ApJ...563...34M,2014ApJ...782...39T,2018ApJ...864..113R}
and (2) the various core-specific scaling relations between the black
hole mass, core size, size of the gravitational sphere of influence
and `missing' light compared to the inwards extrapolation of the
steeper outer light profile \citep[from which the core
`breaks';][]{2007ApJ...662..808L, 2009ApJ...691L.142K,
  2013ARA&A..51..511K, 2013AJ....146..160R,2016Natur.532..340T, 2018ApJ...864..113R}.

From a radius of $r \sim \SI{15}{kpc}$ inwards down to the smallest
resolved scales, the light profile of Holm~15A is almost exponential
(lower panel of Fig.~\ref{fig:holmvlauer}).
\citet{2015ApJ...807..136B} and \citet{2016ApJ...819...50M}
interpreted this as evidence against a large core in
Holm~15A. However, as Fig.~\ref{fig:holmvlauer} shows, Holm~15A 
fits perfectly into the homology of cored BCGs/ETGs.  \citet{Hopkins2009} suggested
that nearly exponential surface brightness profiles on kpc scales
could be ubiquitous among core galaxies as a relic of merger-induced
star-formation bursts in early evolutionary phases prior to the actual
core formation. In their analysis, \citet{Hopkins2009} assumed that
the sphere-of-influence of the black-hole binary is much smaller than
the spatial scale relevant for these ``extra-light'' regions. In fact,
their fits including exponential components often do not well
represent the actual core region. We now know that the sizes of the
cores are almost identical to the sphere-of-influence radii of the
central black holes \citep{2016Natur.532..340T}. The core of Holm~15A
has a size of $3 - \SI{5}{kpc}$ (cf. Fig.~\ref{fig:holmvlauer},
Sec.~\ref{sec:photometry} and also
\citealt{2014ApJ...795L..31L}). Hence, the expected sphere of influence is so large that it interferes with the spatial scale of potential extra-light. The only other galaxy
that seems to be {\it dominated} by a nearly exponential behaviour in its
entire inner region may be NGC~1600 (cf. \citealt{Hopkins2009}). NGC~1600
has a large sphere-of-influence radius of $\SI{1.2}{kpc}$ as
well. There are many processes that influence the final inner
light profile of massive galaxies, like dynamical interactions between
stars and the SMBH binary, early star-formation episodes, AGN feedback
etc. While these processes have been studied individually (in
different levels of detail,
e.g. \citealt{2006ApJ...648..976M,Hopkins2009,2011MNRAS.414..195T,2012MNRAS.422.3081M,2013MNRAS.432.1947M,Choi+2018,2018ApJ...864..113R,2019ApJ...872L..17R}),
we currently lack of simulations that include all these processes in a consistent manner. The black hole binary core scouring process, which is likely dominant in core formation has now been studied in great detail, including the effects of different merger histories on the stellar density profile and stellar orbits in the core  \citep{2018ApJ...864..113R,2019ApJ...872L..17R}.  
Here, we use 
dynamical models based on new spectroscopic observations with
the MUSE IFU\footnote{Based on observations collected at the European
  Organisation for Astronomical Research in the Southern Hemisphere
  under ESO program 099.B-0193(A).} to determine the mass of the central black hole and the distribution of central stellar orbits in Holm~15A. Our 
goal is to shed light on possible formation scenarios for the galaxy's extreme core.

This paper is structured as follows: Section \ref{sec:photometry}
describes the new $i$-band photometry of Holm~15A obtained with the
Fraunhofer Telescope at the Wendelstein Observatory, as well as additional images generated from our MUSE data. Section \ref{sec:obs} details the MUSE spectroscopy and stellar kinematics derived from them. The dynamical models and results based on the photometry and kinematics
are presented in Section \ref{sec:modeling}. In Section
\ref{sec:discussion} we discuss these results and their implications,
in particular in view of predictions from N-body simulations. We
summarize our conclusions about Holm~15A in Section\ref{sec:conclusion}.

We use the Planck $\Lambda$CDM \citep{Planck2018} cosmological model, $H_{0} = 67.4$, $\Omega_{M} = 0.315$. The redshift of Holm~15A, z = 0.055, then corresponds to a luminosity distance of 
$D_{L} = \SI{252.8}{Mpc}$ and an angular diameter distance of $D_{A} = \SI{227.2}{Mpc}$ ($1\arcsec = \SI{1.10}{kpc}$).

\section{PHOTOMETRY}
\label{sec:photometry}
We used two image sources for our photometric analysis
of Holm~15A. 
The first is an $i$-band image obtained with the Fraunhofer Telescope
at the Wendelstein observatory using the Wendelstein Wide Field Imager
\citep[WWFI,][]{2014ExA....38..213K}.
While a $g'$-band image was also available, the
$i$-band image had significantly better seeing (Moffat FWHM from fits to
multiple stars = $0\farcs86$ versus $1\farcs8$ for the $g'$-band
image). 
The isophote analysis of this image is the basis for the 3D deprojection that we use to constrain the
dynamical models (Sec. ~\ref{sec:wendelreduction}). We also used this image to analyse the core region and estimate the "missing light" in
the center of 
Holm~15A (Sec.~\ref{sec:missinglight}).

The second source is an image created from the MUSE data cube, which we used to analyse Holm~15A for the presence of dust or color gradients which could potentially affect the deprojection (Sec.~\ref{sec:musephotometry}, also cf. Sec.~\ref{sec:spectra} for the spectroscopic analysis).

\subsection{Wendelstein image: reduction and PSF-deconvolved light profile}
\label{sec:wendelreduction}
Holm~15A is part of the sample of 170 local BCGs that were observed by \citet{Kluge2019} with the Wendelstein Wide Field Imager. The light profiles derived for these BCGs provide a unique photometric data base, reaching down to an unprecedented deep limiting surface brightness of $\sim 30 \, \mathrm{mag/arcsec^2}$ in the $g'$-band \citep[][ cf. Figure \ref{fig:holmvlauer}]{Kluge2019}.
The data cover a field of  $49 \arcmin{} \times{} 52\arcmin{}$ (pixel size $0\farcs2$/pixel) around Holm~15A, which corresponds to a projected area of roughly $\SI{10}{Mpc^2}$. 
The radial surface brightness profile was measured by fitting ellipses to the galaxy's isophotes, while allowing for higher order deviations from perfect ellipses, using the code from \citet{Bender&Mollendhoff1987}. 
To increase the spatial resolution in the inner parts of the galaxy, the central $\sim 1 \arcmin \times 1 \arcmin$ of the image has been point-spread function (PSF) deconvolved using 40 iterations of the Richardson-Lucy method \citep{Lucy1974}. 
The 2D-convolution is performed on images regenerated from the previously performed isophote analysis. 
The radial light profile from this PSF-deconvolution is the basis of our 3D deprojection that we use to constrain the dynamical models of Holm~15A. 
A detailed description of the observations and
data reduction can be found in \citet{Kluge2019}.

\subsection{Core radius and missing light of Holm~15A}
\label{sec:missinglight}
The core radii of massive galaxies are typically described by either the core-break radius $r_{b}$ of a ``Nuker''- \citep{1995AJ....110.2622L} or 
core-S\'ersic profile \citep{2003AAS...20311620G, 2004AJ....127.1917T}, or by 
the `cusp-radius' $r_{\gamma}$, the radius where $d\log{I}/d\log{r} = -1/2$. The cusp radius only requires that a galaxy's light profile becomes shallow in the central parts. This is clearly the case in Holm~15A and the cusp radius is well defined: $r_{\gamma}$ = $3\farcs7 \pm 0\farcs10$ ($4.11 \pm \SI{0.11}{kpc}$). The semi-major axis length of the corresponding isophote is
$a_{\gamma} =  4\farcs1 \pm 0\farcs10$, consistent with  \cite{2014ApJ...795L..31L}. In contrast, the concept of a core-break radius implies -- in addition to central shallowness -- a distinct change of the light profile from its behaviour outside of $r_b$ to a different behaviour interior to $r_b$. As we will discuss here, the light profile of Holm~15A does not exhibit a clear and distinct change but continously flattens to the smallest observed radii.

The surface brightness distribution of Holm~15A out to $r < 200\arcsec$ (or $\mu_i < 26 \mathrm{mag/arcsec^2}$) can be represented fairly well by the sum of two
S\'ersic functions, where the inner component is nearly exponential with S\'ersic index $n_1 = 1.26$ and $r_{e,1} = 15.81 \, \mathrm{kpc}$ and the outer component follows roughly
a de-Vaucouleurs profile with $n_2 = 4.21$ and $r_{e,2} = 208.1 \, \mathrm{kpc}$ \citep{Kluge2019}.
A more complex model composed as the sum of a core-S\'ersic plus a S\'ersic function improves the fit in the core region slightly. The break radius of this model, $r_b = 8\farcs96$ (cf. model cSS in Tab.~\ref{tab:1dcoretab} of App.~\ref{sec:1dphotometry}) is roughly consistent with the radius of maximum curvature of the observed light profile. However, the S\'ersic parameters of the core-S\'ersic component are very different from the inner S\'ersic component of the model by \citet{Kluge2019} quoted above. The ``steep'' S\'ersic index $n_1 = 5.24$ together with the fact that $r_{e,1} < r_b$ undermine the intended meaning of $r_b$ as a "break radius" and of $n_1$ and $r_{e,1}$ as the local S\'ersic approximation to the light outside of the core. Indeed, the corresponding S\'ersic part of the model does not trace the observed light profile anywhere in the inner regions of the galaxy.

\begin{figure}
\centering
 \includegraphics[width=1\columnwidth]{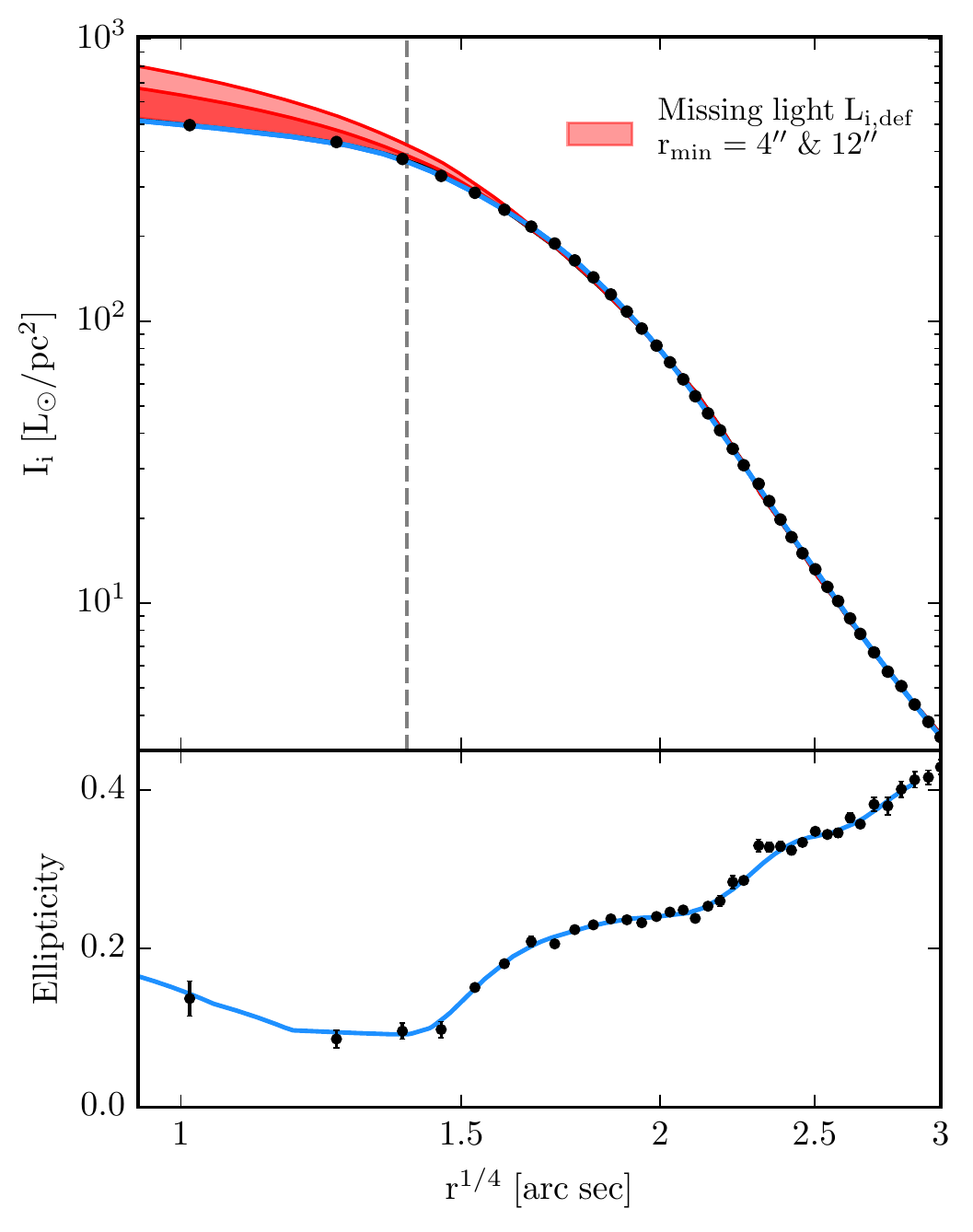}
   \caption{Top: deconvolved i-band light profile of Holm~15A (corrected for extinction and cosmological dimming, black dots) and inwards extrapolation of outer S\'ersic components
   from multi-component (core-)S\'ersic models 
   to the light profile from large radii ($r_{max} \sim 200 \arcsec$) to inner radii of $r_{min} = 4 \arcsec$ and $r_{min} = 12 \arcsec$ (red lines). Red areas indicate  
   the missing light relative to Holm~15A's depleted, shallow core for both models. 
   Bottom: Ellipticity from ellipse fits to the isophotes of Holm~15A.
   Blue lines indicate the projection of our 3D deprojection of the 2D Wendelstein image.
}
   \label{fig:epsandPA}
\end{figure}

To investigate this a little further, we also tried an alternative fitting approach where we separate the determination of the core parameters from the two S\'ersic components:  We start by fitting the sum of two (coreless) S\'ersic components to the surface brightness profile outside of the core, i.e. outside of a minimum radius $r_\mathrm{min}$. Then, in the second step, we repeat the fit, now including also the data inside $r_\mathrm{min}$ but now we only vary the core parameters in the fit, while holding the inner and outer  S\'ersic components $n_1$,$r_{e,1}$ and $n_2$, $r_{e,2}$, and $\mu_{e,2}$ fixed. In this way we determine the S\'ersic parameters before the core parameters and force the S\'ersic components to approximate the light profile outside of $r_\mathrm{min}$. We tried a range of different $r_\mathrm{min}$. Below $r_{min} < r_\gamma \sim 4\arcsec$ (i.e. inside the core) the inner components $n_1$ and $r_{e,1}$ are too much affected by the core region itself. Above $r_\mathrm{min} = 12\arcsec$, the light profile is already so steep that we are far outside the core and the models, even after fitting the core parameters, do not provide good fits anymore. For $r_\mathrm{min} = 4\arcsec-12\arcsec$, these two-step fits represent the data very well. Moreover, in all the two-step fits we found $r_b < r_e$, and $r_{b} \sim r_\gamma$, as expected (cf. models cSS($r_\mathrm{min}=4$) and cSS($r_\mathrm{min}=12$) in Tab.~\ref{tab:1dcoretab}). The S\'ersic components approach the  S\'ersic+S\'ersic model of \citet{Kluge2019} in the limit of small $r_\mathrm{min}$. However, the fits did not converge to a stable set of parameters. We found both the S\'ersic index $n_1$ of the inner component and $r_b$ to systematically increase with $r_\mathrm{min}$ (cf. Tab.~\ref{tab:1dcoretab} and Fig.~\ref{fig:epsandPA}). 

All these results led us to conclude that the galaxy does not exhibit a clear break radius inside of which the light profile follows a power law and outside of which it can be characterised by a single local S\'ersic index $n$ over a range that is more extended than a few arcseconds. Fitting the inner parts of the 1D light profile of Holm~15A with a Nuker profile confirmed this finding. Again, we could not derive a stable break radius and $r_b$ turned out to be a monotonic function of the maximum radius out to which we extended the fit (we tried $r_\mathrm{max} = 10-70 \, \arcsec$; cf. Tab.~\ref{tab:1dcoretab}). 

Finally, we also performed a 2D- multi-component fit to the entire $i$-band Wendelstein image of Holm~15A using IMFIT (\citealt{2015ApJ...799..226E}; see Appendix~\ref{sec:irafanalysis}). This yielded a stable set of core parameters. However, in the 2D-analysis, allowing for a broken inner profile with a power-law core did not improve the fit significantly over a central, pure S\'ersic component with $n \sim 1$ and $r_{b} = 2\farcs57$. 

Holm~15A evidently continues the homology of cores observed in less extreme ellipticals in the sense of having a faint center with a shallow surface brightness profile (cf. Fig.~\ref{fig:holmvlauer}). But, as our attempts of identifying a clear break radius have shown, the core region in Holm~15A is not as sharply separated from the outer parts of the galaxy as it is in other core galaxies with a more prominent break in the light profile. 
Because of this, even though both $r_{b}$ and $r_{\gamma}$ have been shown to follow tight scaling relations with $M_{BH}$ in other core galaxies \citep[e.g.][]{2007ApJ...662..808L, 2016Natur.532..340T}, we will only consider the cusp radius of Holm~15A in the rest of the paper. 

The shallowness of the inner light profile still allows the estimation of the amount of "missing light". From the above described models cSS($r_\mathrm{min}=4$) and cSS($r_\mathrm{min}=12$) (see Tab.~\ref{tab:1dcoretab}) we find $L_{i, def} = (2.75 \pm 2.22) \times 10^{10} L_{i, \odot}$, which we will later use in Section \ref{sec:scaling} to estimate the mass of stars ejected from the center via core scouring. The estimated missing light is illustrated in Fig.~\ref{fig:epsandPA}. 

\label{sec:musephotometry}

\subsection{MUSE images: no evidence for dust or color gradients}
\label{sec:musephotometry}
To investigate whether dust extinction might distort the
isophotes, and to check for color
gradients indicative of a change in the stellar populations, we also generated images from the MUSE data cube. 
This has two advantages. First,
the MUSE observations have (slightly) better seeing than the Wendelstein
$i$-band image: in the ``red'' image (see below for definition), we
measured FWHM = 0\farcs72 from the two point sources in the image.
Second, when collapsing the data cube we can choose wavelength ranges
that explicitly exclude emission, which is important because we do
detect regions of line emission within Holm~15A (see below).

\begin{figure*}
\centering
 \includegraphics[width=1.8\columnwidth]{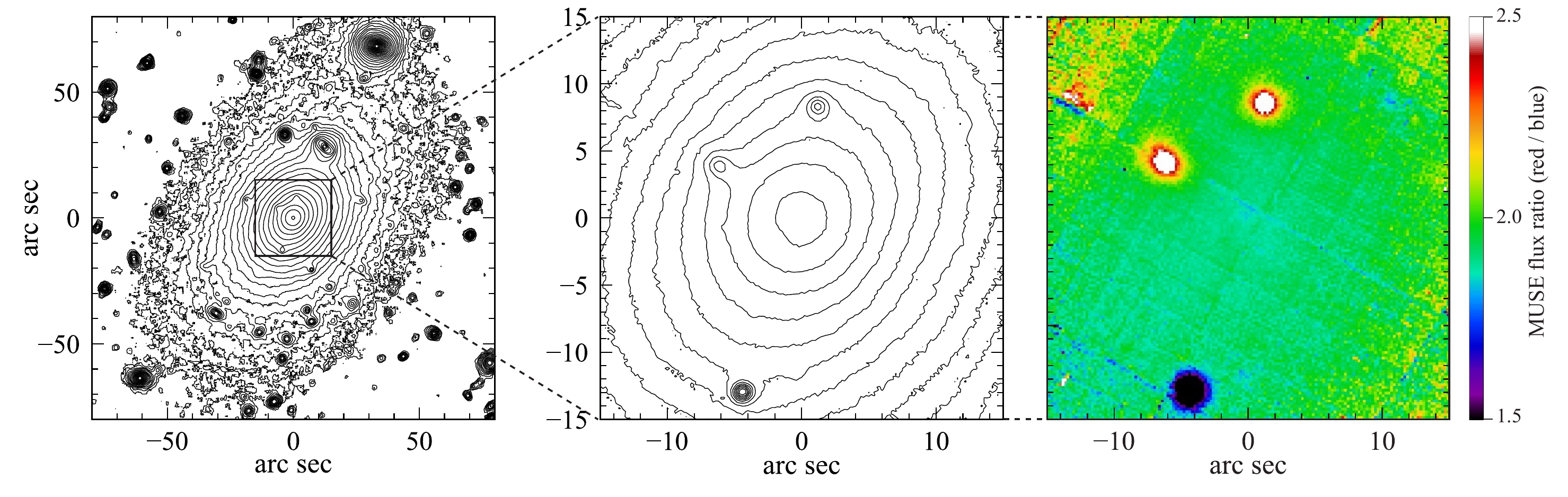}
   \caption{Holm~15A isophotes and central color map. Left: Logarithmically
scaled isophotes for our Wendelstein $i$-band image (median-smoothed
with an 11-pixel-wide box). Middle: Isophotes for the MUSE red image
(extracted from data cube using 7300--8500 \AA). Right: Color map from
ratio of MUSE blue (4750--5500 \AA) and red images.
No evidence for dust lanes or a color gradient in the central region of
the galaxy can be seen.
}
   \label{fig:contourscolors}
\end{figure*}

We use the spectral region 7300--8500 \AA{}  to create a largely emission-line-free
``red'' image and the
spectral region 4750--5500 \AA{} for its ``blue'' counterpart. 
The ratio of the blue and red MUSE images is shown in
the right-hand panel of Figure \ref{fig:contourscolors} and it shows
\textit{no} evidence for either dust lanes or significant color
gradients.

\subsection{3D deprojection}

\label{sec:depro}
In order to constrain the distribution of stars in  our dynamical model of Holm~15A (see Section \ref{sec:modeling}),
we create a 3D deprojection of 
the luminosity density from our deconvolved 2D Wendelstein image. The algorithm that we use to achieve this enables us 
to find a 3D non-parametric axisymmetric luminosity density distribution $\nu(\bm{r})$ consistent with the 2D
input surface brightness distribution 
and an assumed inclination angle $i$.
As can be seen in Figure \ref{fig:epsandPA}, Holm~15A is for the most part relatively round, 
but flattens significantly to an ellipticity $\varepsilon \sim 0.4$ 
at radii $\gtrsim 100 \arcsec$.
In the axisymmetric case, this limits possible viewing angles to be close to edge on, which is why we assume $i = \ang{90}$.
 The algorithm utilizes a penalized log-likelihood function 
and is detailed in \citet{1999MNRAS.302..530M}. As Figure \ref{fig:epsandPA} shows, the resulting axisymmetric luminosity density distribution reproduces the relevant observed photometric features almost perfectly. 

\section{MUSE spectroscopy: stellar kinematics of Holm~15A}
\label{sec:obs}
\subsection{MUSE observations and data reduction}

\label{sec:muse} 
We obtained wide-field spectroscopic data of Holm~15A from the
Multi-Unit Spectroscopic Explorer (MUSE) at the \textit{Very Large Telescope}
at Paranal on 2017 November 16 and 2018 August 10. At z = 0.055
MUSE covers several important absorption features such as H$\beta$,
the Mgb region, NaI, several Fe absorption features and the Ca II
triplet.  

Our observations were carried out over the course of two
nights and consist of three observational blocks of two dithered
\SI{1200}{s} exposures of Holm~15A plus one \SI{300}{s}-long exposure of the
sky, inbetween each. All observations, including the sky-field offset, cover an
approximately $\SI{1}{\arcmin} \times \SI{1}{\arcmin}$ FOV composed of
24 combined integral field units (IFUs).  \\We performed the data
reduction using version 2.8.5 of the standard Esoreflex MUSE pipeline
supplied by ESO \citep{2013A&A...559A..96F}. The pipeline runs several
recipes on both exposures such as flat-field and wavelength
calibrations and returns a combined data cube, covering the optical
domain from about $\SI{4800}{\angstrom}$ to
$\SI{9400}{\angstrom}$ with a spectral resolution of $\SI{1.25}{\angstrom}$. We sampled
the cube in spaxels of $0\farcs4 \times 0\farcs4$, which at the redshift of the galaxy (z = 0.055) corresponds to
approximately $\SI{400}{pc} \times \SI{400}{pc}$ per pixel. As previously mentioned,
we measure a PSF with FWHM = $0\farcs71$ for the MUSE image.

Sky emissions were removed separately from all galaxy exposures using
the sky-field from offset sky-exposures, taking into account the
  instrumental line spread function for each IFU.

\subsection{Treament of spectra and derivation of (parametric) stellar
  kinematics}
\label{sec:spectra}
For our study of Holm~15A, we initially used the MUSE absorption
spectra to derive spatially resolved, 2D stellar kinematics
parameterized by the rotational velocity $v_{rot}$, velocity
dispersion $\sigma$ and higher-order Gauss-Hermite coefficients
$h_3$ and $h_4$ of the line-of-sight velocity distribution (LOSVD).  For the dynamical
  modelling, we use non-parametric LOSVDs that were derived following
  a set of equivalent steps (see Sec.~\ref{sec:wingfit}).

  To achieve a balance between a precise measure of the kinematics in the core and an
  overall high spatial resolution we aim for a target
  S/N of at least $\sim 50$ per pixel in
  each spectrum. To achieve this, we spatially bin the 
  data cube using the Voronoi tessellation method of
  \citet{2003MNRAS.342..345C}. Pixels belonging to foreground sources
  such as galaxies or AGN are removed from the data before binning.

  At the center of the galaxy ($r \leq 5 \, \mathrm{kpc}$) the spatial
  resolution of the Voronoi bins turns out to be $0.4''$ - $0.8''$
  (roughly $400 - \SI{800}{pc}$) for a S/N $\sim 50$.
  We here define the radius of the gravitational sphere of influence (SOI)
  of the black hole as the radius where the enclosed mass $M(\leq r_{SOI}) \equiv M_{BH}$. By integrating the
  deprojected 3D luminosity density and assuming a
   range of plausible stellar mass-to-light ratios, between $\Upsilon_{*} = 4$ and $6$, we estimated the enclosed mass of the galaxy.
  For the lowest expected black hole mass for a galaxy of this mass and 
  velocity dispersion, $M_{BH} \sim 3\times 10^9 \si{M_{\odot}}$ \citep[using
  the mean expected values from the $M_{BH}-\sigma,M_{Bu}$ scaling relations for ETGs from][]{mcConnell&ma2013, 2016ApJ...818...47S}), the enclosed stellar mass equals $M_{BH}$ at $r_{SOI} \sim 1\farcs6$. Since our PSF and spatial binning resolution are both on the order $0\farcs8$ we ensure that we can resolve the expected sphere of influence (SOI) with a diameter of $2 \times 1\farcs6 = 3\farcs2$ 
  by a factor $\geq 4$. However, the extreme core properties of Holm~15A actually
  point to a SMBH with $M_{BH} \sim 10^{11} M_{\odot}$ (based on $M_{BH}-r_{\gamma}$ scaling relations from \citet{Lauer2007b} and
\citet{2016Natur.532..340T}), whose SOI radius would be roughly $r_\mathrm{SOI} \sim 4-5 \, \arcsec$ -- a factor $>10$ above our resolution limit. If
  the dark matter halo is included in the modeling, this resolution is
  sufficient for a robust black hole mass determination
  \citep{2013AJ....146...45R}.  
  
  In total, we obtain 421 spatial bins,
  of which 145 bins are located inside the central $5''$. For the purpose of our subsequent dynamical
modeling of the galaxy we divided the spatial bins of our MUSE FOV
into four quadrants, q1-4 in such a way that quadrant membership is
determined by which side of the major and minor axes the center of
each bin is located on

\begin{figure}
  \includegraphics[width=0.9\columnwidth]{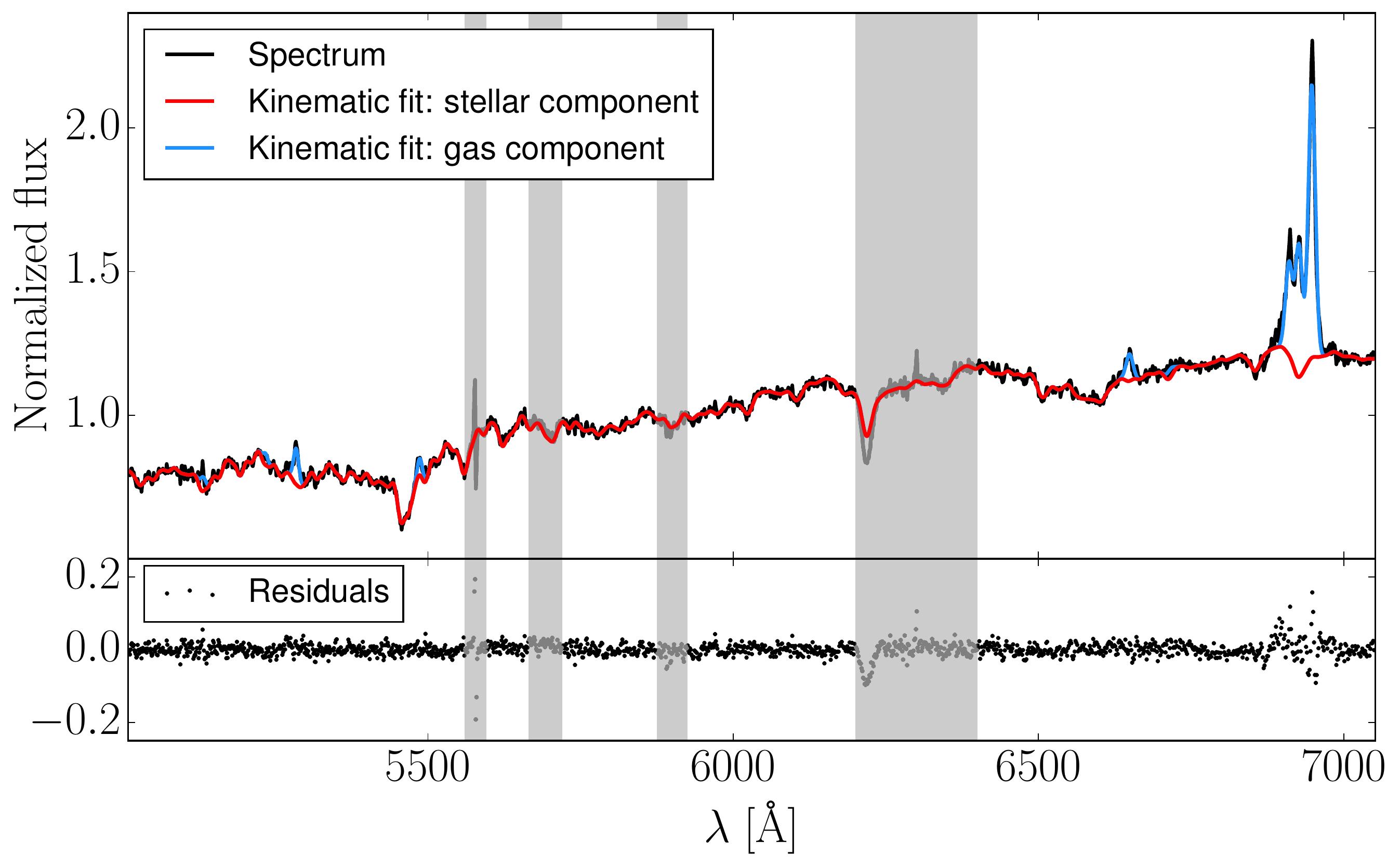}
  \caption{Stellar kinematic fit with pPXF (red) to a normalized
    spectrum of Holm~15A (black) with corresponding residuals
    (black points, lower panel). Emission lines from ionized gas are
     fit simultaneously (blue).
    Spectral regions masked during the fit are shown as gray shaded areas.}
   \label{fig:specs}
  \end{figure}

Parametric LOSVDs for each bin were obtained by fitting the stellar absorption
lines of the galaxy with Penalized Pixel-Fitting
\citep[pPXF,][]{2017MNRAS.466..798C} implemented in Python 2.7. PPXF
convolves a weighted sum of template stellar spectra, in this case the
MILES library \citep{2006MNRAS.371..703S} with a
Gauss-Hermite LOSVD in order to fit the absorption features. Optionally, 
emission-line features of ionized gas are fit simultaneously, with a separate set of templates and LOSVDs. 
Figure \ref{fig:specs} shows an example of a
(parametric) kinematic fit to the
spectral features of Holm~15A with pPXF for a bin located roughly
$0\farcs5$ from the center of the galaxy (best fit to stellar component:
 $v_{rot} = -1.59 \pm \SI{8.04}{km/s}$ 
relative to the systemic velocity of the galaxy, $\sigma = 342 \pm \SI{9.71}{km/s}$,
$h_3 = 0.025 \pm 0.015$, $h_{4}= 0.062 \pm 0.018$).

\begin{figure}
 \includegraphics[width=1\columnwidth]{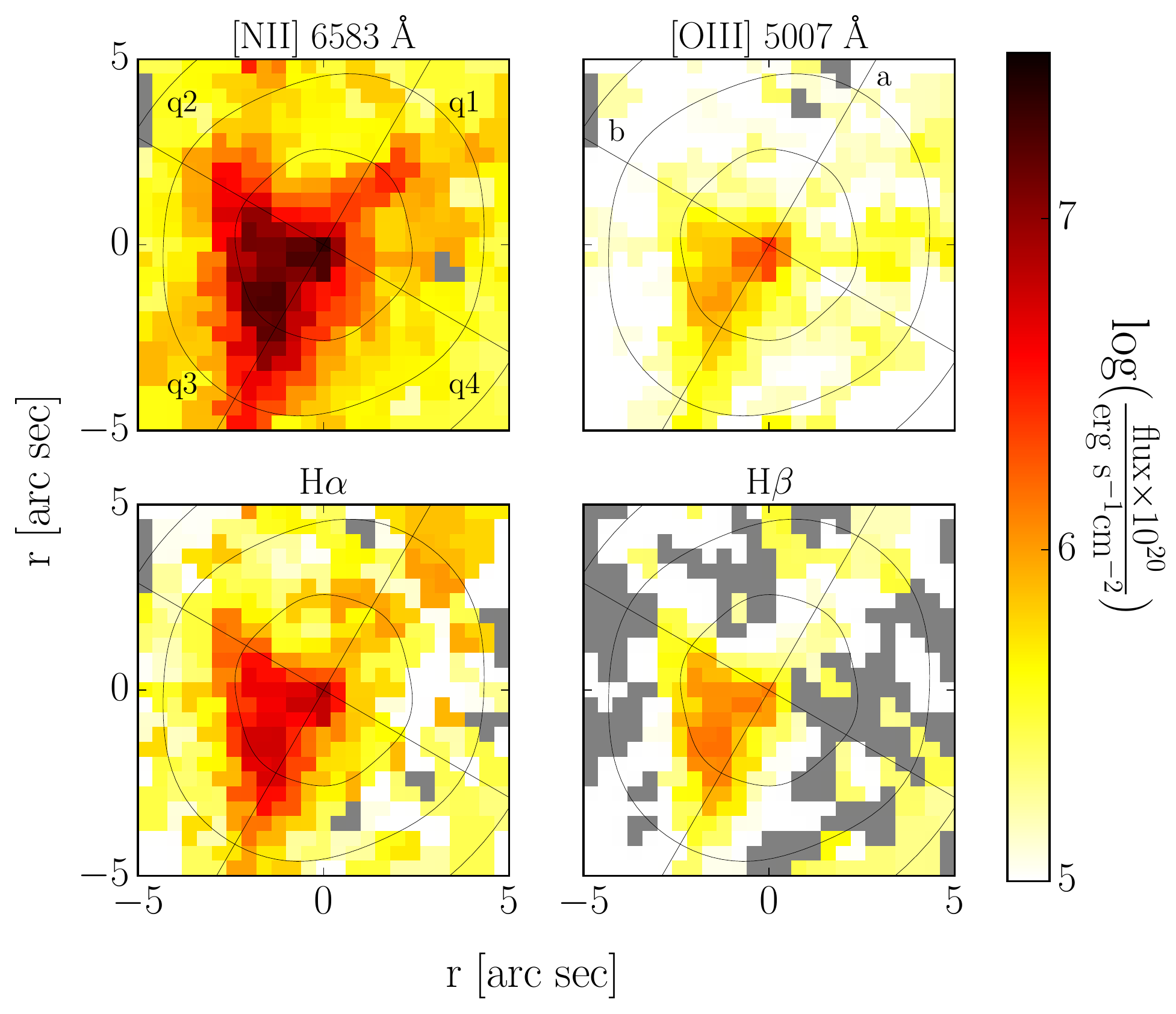}
  \caption{Logarithmic flux of emission lines [NII], [OIII], H$\alpha$ and H$\beta$
	  from ionized gas located within central regions
	  of the galaxy. Grey areas indicate bins for which no meaningful emission line-fit 
	  could be derived. Photometric i-band isophotes are shown in black. axes $a$ and $b$ (black lines) correspond to the major-
     and minor axes of the galaxy respectively.
	The center of the galaxy coincides with the peak of the emission line flux.}
  \label{fig:gasmap}
\end{figure}

Several bins within the central $\SI{5}{kpc}$ of the galaxy -- primarily in the southeastern regions -- region contain emission lines
from ionized gas, most notably H$\alpha$, H$\beta$ , [OIII] $\SI{5007}{\angstrom}$, [NI] $\SI{5199}{\angstrom}$ and [NII] $\SI{6583}{\angstrom}$ (cf.
Figure \ref{fig:specs}), which we fitted with the emission line
fitting routine of pPXF, though we do not consider their kinematics in this study. Figure \ref{fig:gasmap} shows the measured emission line flux for H$\alpha$, H$\beta$, [OIII] and [NII]. The average flux ratios $\log(\mathrm{[OIII]/H\beta}) = 0.09 \pm 0.26$ and
$\log(\mathrm{[NII]/H\alpha}) = 0.48 \pm 0.12$ of emission lines with
S/N > 3 are associated with LINER-type emission \citep{2003MNRAS.346.1055K}, which is quite typical for cool-core clusters. Of the $\sim 100$ brightest X-ray clusters, Abell 85's cool core has the 14th strongest cooling flow \citep{Chen2007}. The spatial extent of this LINER-type emission ($\sim 4$--\SI{5}{kpc}) suggests it could be related to ionized cooling-flow filaments 
\citep[e.g][]{2008MNRAS.386L..72F, 2009MNRAS.392.1475F, Ogrean2010}.
This was already previously noted by \cite{2010ApJ...721.1262M}, who found that it coincided with a similarly extended region of X-ray emission associated with cooling flows. 

By contaminating some absorption features such as H$\beta$, the
gas emission increase the uncertainties of the kinematic fits
in some bins.  As we will show in section
\ref{sec:modeling}, this contamination of mostly central spectra 
slightly increases the uncertainty of
$M_{BH}$, but has little impact on the global stellar
mass-to-light ratio $\Upsilon_{*}$ and the shape of the dark matter
halo.
\\At redshift z = 0.055, the strong oxygen $\SI{5577}{\angstrom}$
sky emission line lies on top of the $\SI{5270}{\angstrom}$
Fe-feature. Because this line is difficult to remove, the Esoreflex
sky subtraction left strong residuals in this region, effectively
rendering it unusable for fitting. We noted a few additional
systematic residuals which may be related to sky subtraction or telluric correction 
issues as
well. In order to minimse possible systematics in the LOSVDs, we
  defined a single mask that we used for all spectra throughout the
  entire galaxy. We consistently mask all wavelength regions that are
  possibly affected by any systematic
  issues.
    
  We performed our kinematic fits over the spectral
  interval between 5010 and \SI{7050}{\angstrom}.
  Including spectral regions bluer than $\SI{5010}{\angstrom}$
  resulted in lower-quality fits and a constant bias in $h_3$, indicative
  of template mismatch.  Spectral regions redder than
  $\SI{7050}{\angstrom}$ were badly affected by sky lines and were
  therefore omitted.  In particular, we could not derive meaningful
  kinematics in the [Ca II] triplet region.  
  
  We also used a
  6th order multiplicative polynomial and an additive constant in the
  fit.  The former allows for the correction of errors in the flux
  calibration, while the latter is typically used to correct over- or
  underestimations of the continuum during sky correction. We also
  made use of the sigma clipping and bias-factor options.  The value
  of the bias factor -- 0.2 in our case -- was determined from testing
  pPXF on Monte Carlo simulations of model spectra.  
  \\A subset of stellar template spectra for the fit was selected as follows: We fitted
  a mean spectrum of all bins of the galaxy with the full set of 985
  MILES library templates.  All binned spectra were corrected for the systematic 
  velocity of the galaxy, as well as their respective rotational velocities. All spectra were normalized to one before averaging. We set both the third-order Gauss-Hermite coefficient $h_{3}$ and the 
additive constant to zero in order to avoid template mismatch (which can
result in biases in these parameters). With these restrictions pPXF assigned non-zero
  weights exclusively to a set of 16 templates with a wide variety of luminosity classes 
  but limited to spectral types G, K and M, in good
  agreement with the uniformly red color of the galaxy
  (ec.~\ref{sec:photometry}). We used this subset of
  stars from the MILES library as templates
  for fitting the galaxy's absorption features in all
  Voronoi bins.

\begin{figure}
 \includegraphics[width=1.\columnwidth]{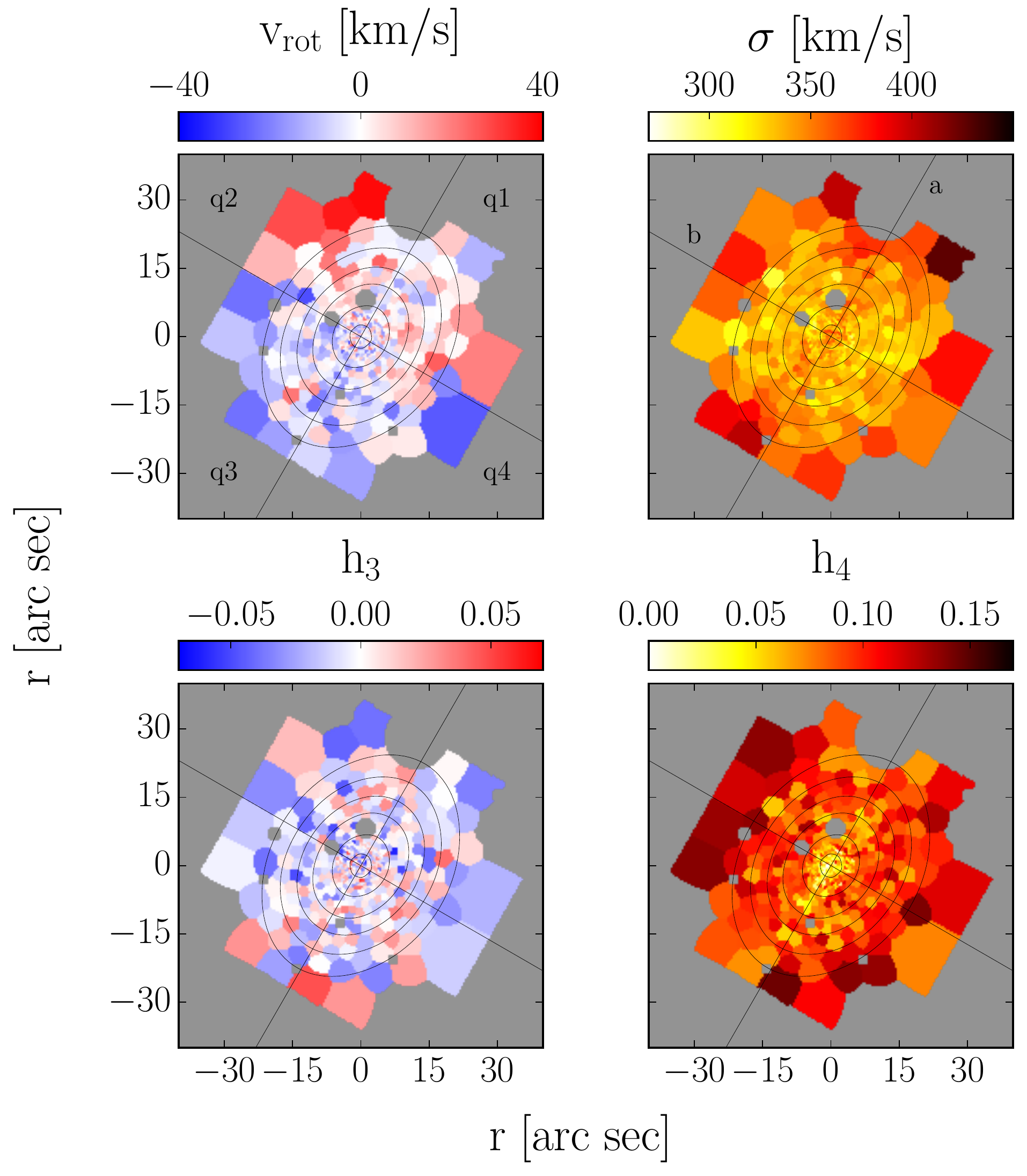}
 \centering
 \caption{From top to bottom, left to right: kinematic maps of the
   rotational velocity $v_{rot}$, velocity dispersion $\sigma$ and the
   higher-order Gauss-Hermite coefficients $h_3$ and $h_{4}$ over the MUSE
   FOV. The systematic velocity of the galaxy has
   been subtracted in the
   kinematic map of $v_{rot}$.  Ellipse fits to i-band isophotes are
   drawn in black; axes $a$ and $b$ (black lines) correspond to the major-
   and minor axes of the galaxy respectively.}
  \label{fig:kinmaps}
\end{figure}

\begin{figure}
\includegraphics[width=1.\columnwidth]{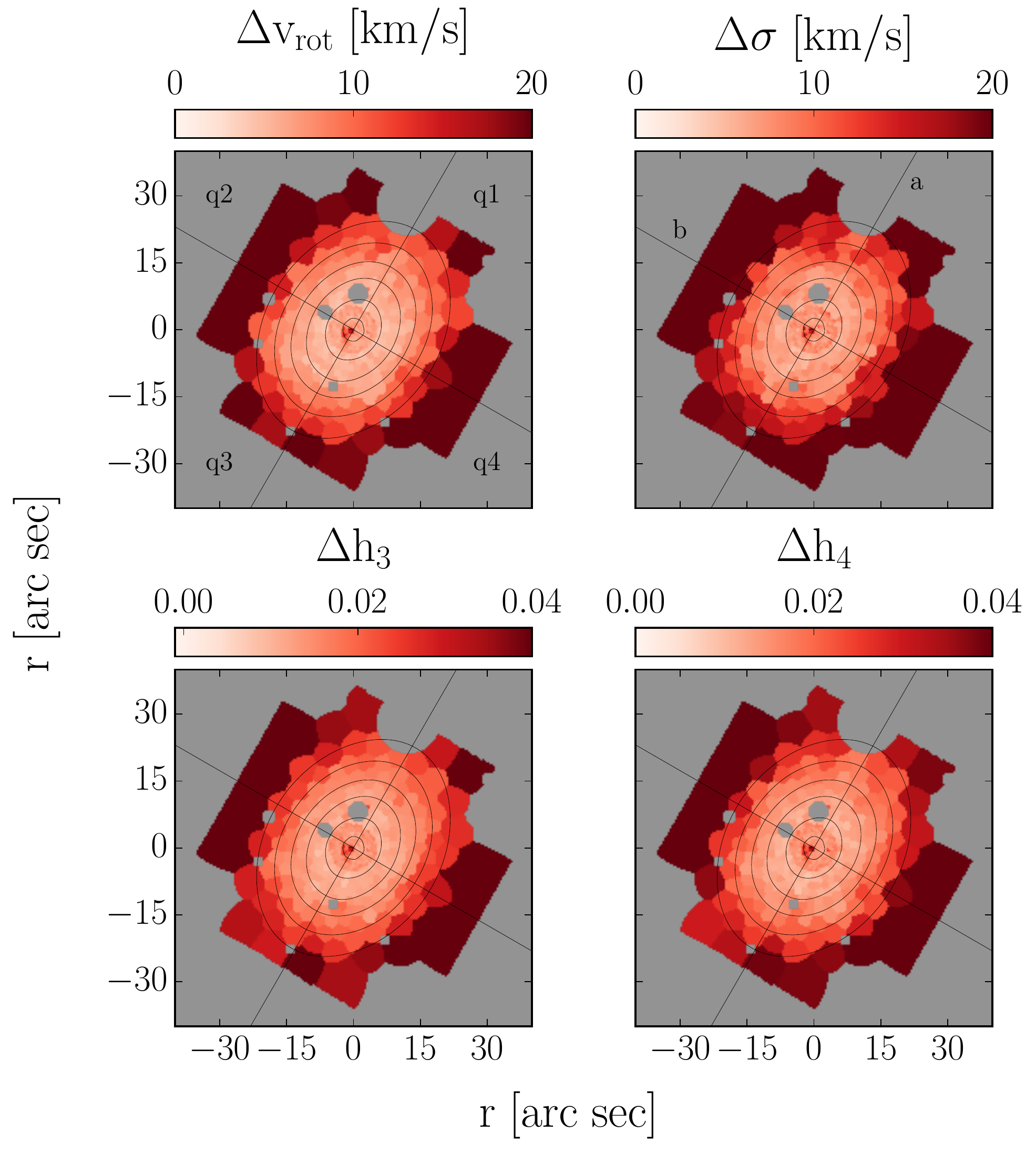}
\caption{Maps of statistical uncertainties corresponding to the
  parameters of the kinematic maps of 
  Figure \ref{fig:kinmaps}.}
  \label{fig:errmaps}
\end{figure}

The parameterized kinematics in the interval between 5010 and \SI{7050}{\angstrom} over the MUSE
FOV are shown in Figure
\ref{fig:kinmaps}. As can be seen in the figure, we
measure a weak rotation signal of less
than $\SI{40}{km/s}$, which is only faintly reciprocated in $h_3$ --
the rotation is likely too weak for an anti-correlated signal in this
parameter to be detectable.  The velocity
dispersion $\sigma$ peaks in the central regions ($r < \SI{2}{kpc}$)
at $\sim \SI{350}{km/s}$, stays somewhat constant at $\sim$
$\SI{330}{km/s}$ throughout most of the FOV and finally starts to rise
again at the edges of the MUSE FOV up to
$\gtrsim \SI{370}{km/s}$. Our measured velocity dispersions are similar to those of \citet{2014MNRAS.443..485F}. Our $h_{4}$
kinematic profile starts out at $\sim 0.07$ within $\SI{2}{kpc}$ and
rises to $\gtrsim 0.1$ towards the edges of the
FOV. In Appendix \ref{sec:kincompare} we compare the kinematics
of Holm~15A to those of massive ETGs from the MASSIVE survey.
 The corresponding statistical uncertainties are shown in
Figure \ref{fig:errmaps}. Uncertainties were determined from Monte Carlo simulations
  on model spectra of the galaxy, i.e. re-fitting best-fit spectral
  models with 100 different noise realizations, the noise being drawn from a
Gaussian distribution with a
  dispersion corresponding to the local S/N, which is measured
  directly from each spectrum. We note that the distribution of uncertainties is 
  spatially asymmetric
  between central bins across quadrants
  -- central kinematics in q3 have overall larger uncertainties than those in the other quadrants.
  This is in agreement with the 
  distribution of emission-line flux between quadrants (cf. Figure \ref{fig:gasmap}), i.e. q3 seems to be affected worse by uncertainties
  introduced by gas contamination of absorption features. However, as we will show in Section \ref{sec:modeling}, including q3 in our dynamical
  modeling did not produce any larger systematic offset in our best fit parameters relative to the other quadrants.

\subsection{Non-Parametric LOSVDs}
\label{sec:wingfit}
In our dynamical modeling of Holm~15A we set out to achieve a precise
mass measurement of the galaxy, which makes the parametric
representation of the stellar kinematics in Figure \ref{fig:kinmaps}
problematic: Large values of $\sigma$ and $h_{4}> 0$ over the entire
FOV result in the escape velocity of the galaxy, $v_{esc}$ being practically
infinite everywhere. Since $v_{esc}$
  depends directly on the gravitational potential we try to measure it
  as accurately as possible.

To obtain LOSVDs with more realistic $v_{esc}$, we use our own
kinematic extraction code (Thomas et
al. in prep.) which operates in a similar way as pPXF but
minimizes the $\chi^2$
over all spectral pixels by utilizing a Levenberg-Marquardt algorithm
to fit a template broadened with a non-parametric LOSVD to the absorption features of a galaxy.  

\begin{figure}
      \includegraphics[width=0.7\columnwidth]{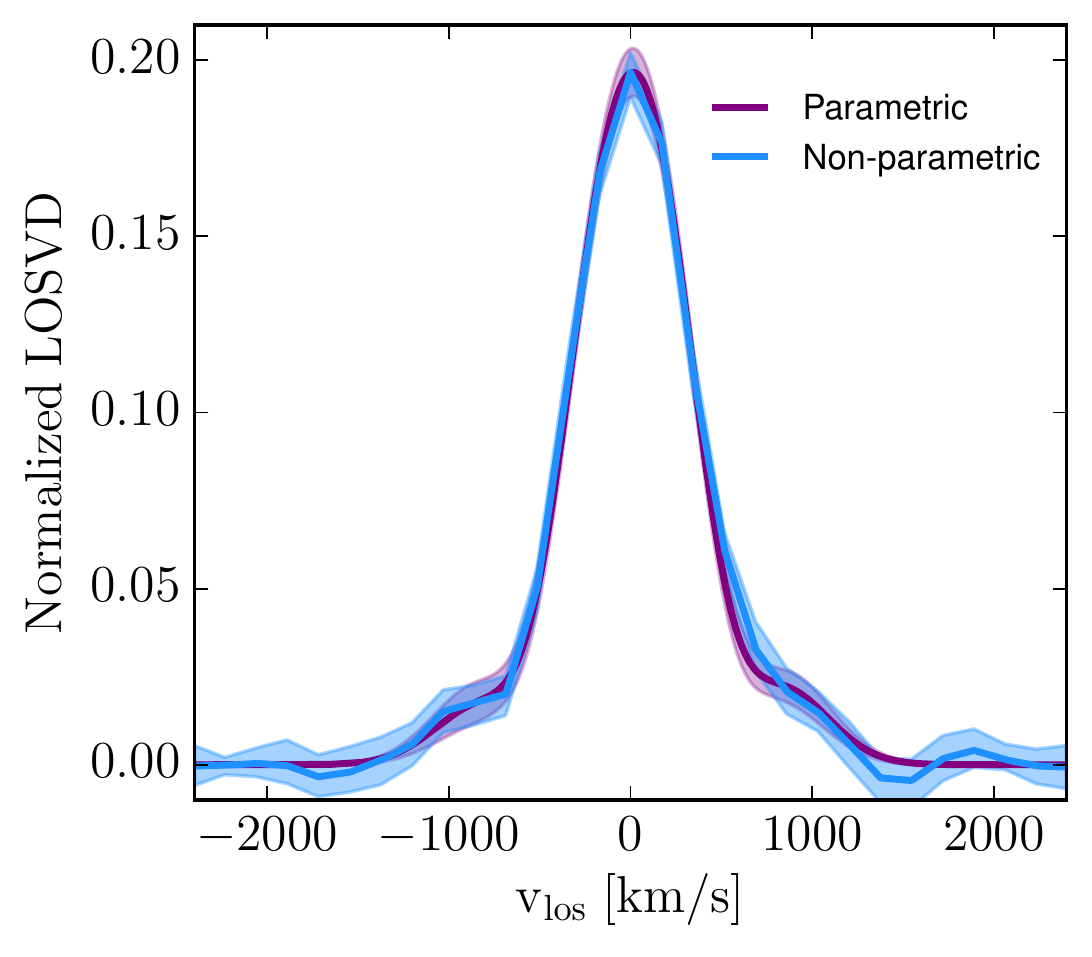}
	\centering
       \caption{Example LOSVDs from the central regions of q4 originating from two different methods: 
       One determined parametrically with pPXF (purple) and the other non-parametrically with our own code
       (blue). The shaded envelopes indicate statistical uncertainties.}
	\label{fig:winglosvds}
\end{figure}

We use the same setup of template stars, additive and multiplicative polynomials
as described above. Emission lines are masked for each 
spectrum individually, according to their respective widths 
(spectral regions within $4 \times \sigma_{gas}$ are masked for each emission line) and positions 
as determined with the pPXF emission line fit.
The non-parametric LOSVDs mainly differ from the parametric
ones in the high-velocity tails, as demonstrated for an example bin of Holm~15A
in Figure~\ref{fig:winglosvds}. 
 While the width of the LOSVD ($\sigma = 338 \pm \SI{9.57}{km/s}$ with our own code and  $\sigma = 328 \pm \SI{10.7}{km/s}$ with pPXF), as well
 as its global shape, are similar
for both methods, the non-parametric LOSVDs provide a more realistic
sampling of the LOSVD and noise at large projected velocities. Therefore,
for our dynamical study of Holm~15A, we use the non-parametric LOSVDs.
Radial profiles comparing both parametric and non-parametric kinematics for all bins in our study
are presented in Appendix \ref{sec:nonParavPara}.

\section{Schwarzschild Dynamical Modeling of Holm~15A}
\label{sec:modeling}
\subsection{Dynamical models}
We dynamically modeled Holm~15A under the assumption of
axisymmetry. The lack of unambiguous, obvious isophotal distortions (see Section
\ref{sec:photometry}) and the overall symmetry of the observed
kinematic profiles (see Section \ref{sec:spectra}) imply that
Holm~15A is generally consistent with an axially symmetric stellar
distribution.  

The dynamical models in this study were constructed using an updated
version of our axisymmetric Schwarzschild orbital superposition code.
We will here only briefly
summarize  the key features of our implementation and refer to previous publications for
more in-depth descriptions \citep{1988ApJ...327...82R,gebhardt03,2004MNRAS.353..391T,siopis09}. 

Schwarzschild dynamical modeling is based on the calculation of
stellar orbital distributions in a fixed gravitational potential as a
solution to the collisionless Boltzmann equation \citep{1979ApJ...232..236S}.  Any orbit can
  be fully described by three integrals of motion: The classical
  integrals $E$ and $L_{z}$ (in the axisymmetric case) plus a
  non-classical integral $I_{3}$ (in most astrophysically relevant
  cases). Sampling values of this set of integrals of motion ($E$,
$L_{z}$, $I_{3}$) allows us to create an orbit library in a given
gravitational potential $\Phi$ whose distribution function
$f(\bm{r}, \bm{v})$ satisfies the collisionless Boltzmann equation.

In order to determine $\Phi$, we assume that the density distribution
of Holm~15A can be described by
\begin{equation}
\rho(r, \theta) = \rho_{\star}(r, \theta) + M_{BH}\delta(r) + \rho_{DM}(r),
\label{eq:fullrho}
\centering
\end{equation}
which we insert into Poisson's equation.  $\rho_{\star}$ is linked to
the three dimensional deprojection $\nu(r, \theta)$ of the observed
i-band surface brightness (cf. Section \ref{sec:photometry}) via the stellar (i-band) mass-to-light
ratio,
$\rho_{\star}(r, \theta) = \Upsilon_{\star} \cdot
\nu(r, \theta)$,
assuming a spatially constant stellar $\Upsilon_{\star}$.  In addition to the
mass of the central black hole $M_{BH}$, the model admits the inclusion of a
dark matter (DM) halo $\rho_{DM}(r)$. Here, we chose a
generalised NFW-halo derived from cosmological N-body simulations
\citep{1996ApJ...462..563N,1996MNRAS.278..488Z}:
\begin{equation}
\rho_{DM}(r) =  \frac{\rho_{0}}{\left(1 + \frac{r}{r_{s}}\right)^{3 - \gamma}\left(\frac{r}{r_{s}}\right)^{\gamma}},
\label{eq:NFW}
\centering
\end{equation}
with 
\begin{equation}
 \rho_{0} = \rho_{10}\left(1 + 10\frac{\si{kpc}}{r_{s}}\right)^{3 - \gamma}\left(10\frac{\si{kpc}}{r_{s}}\right)^{\gamma} ,
 \centering
 \label{eq:rho0}
\end{equation}
where $\rho_{10}$ is the DM density at $\SI{10}{kpc}$, $r_{s}$ the
scale radius of the halo and $\gamma$ the inner slope of the DM
density profile.  

For a given $\Phi$, we sample thousands of representative initial
orbital conditions, implicitly varying all the 
integrals of motion $E$, $L_{z}$ and $I_{3}$, and including individual
orbital phase-space volumes \citep{2004MNRAS.353..391T}. For Holm~15A, we stored LOSVDs in 29
velocity bins adapted to the velocity dispersion of the galaxy,
with one LOSVD associated with each of the 421 spatial bins of our
FOV, meaning our models fitted roughly a total of 3000 velocity bins
per quadrant. 

 We use the NOMAD optimization software
 \citep{AuDe2006, Le2011a, AuHa2017a} to find the set of mass parameters $M_{BH}$,
  $\Upsilon_{\star}$, $\rho_{10}$, $r_{s}$ and $\gamma$ that yields
  the best fit to the observed kinematics.

\subsection{Results}
\label{sec:results}
\begin{figure}
      \includegraphics[width=1\columnwidth]{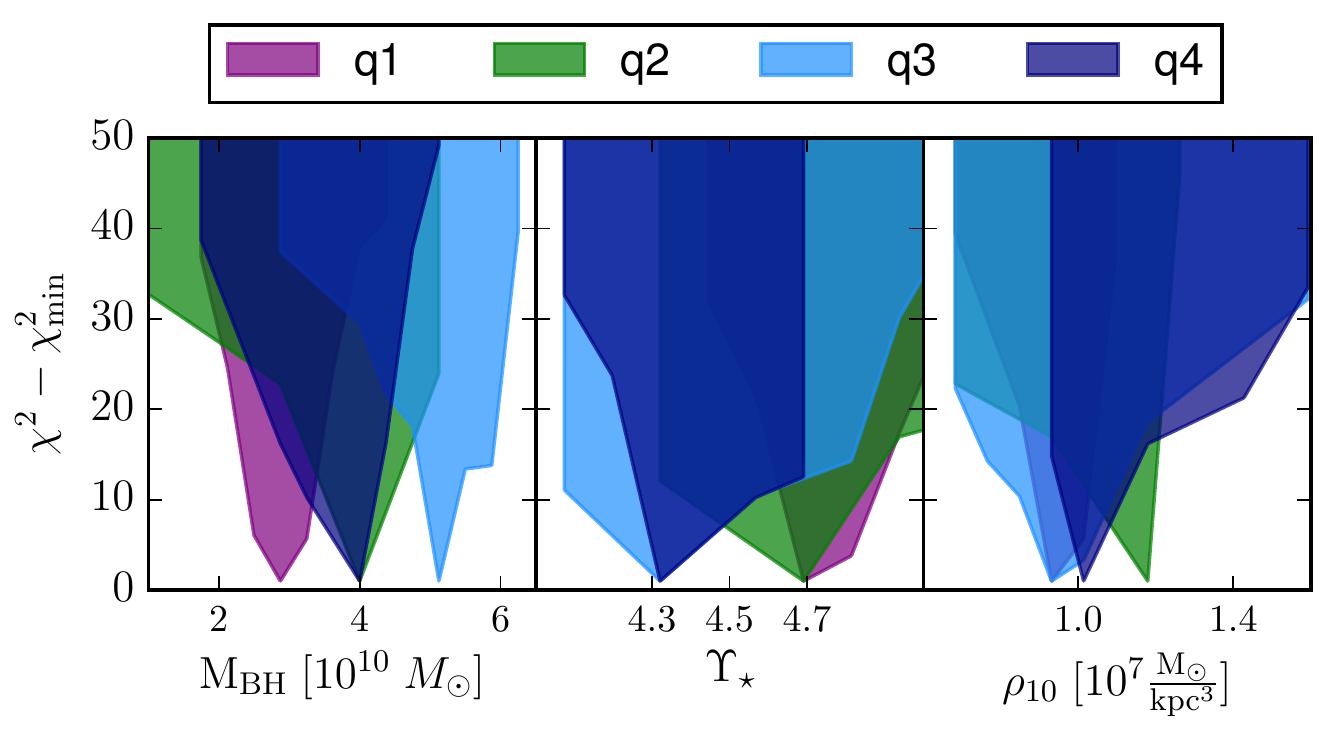}
	\centering
        \caption{From left to right: $\chi^2$
          for our minimization curves of our dynamical modeling for the parameters
          $M_{BH}$, $\Upsilon_{\star}$ and $\rho_{10}$. Each quadrant (q1-4) was modeled separately. The variation between their
          respective $\chi^2$ curves is treated as representative of
          the inherent systematic and statistical uncertainties of each
          measurement.}
	\label{fig:chi}
\end{figure}

\begin{figure}
      
      \includegraphics[width=1\columnwidth]{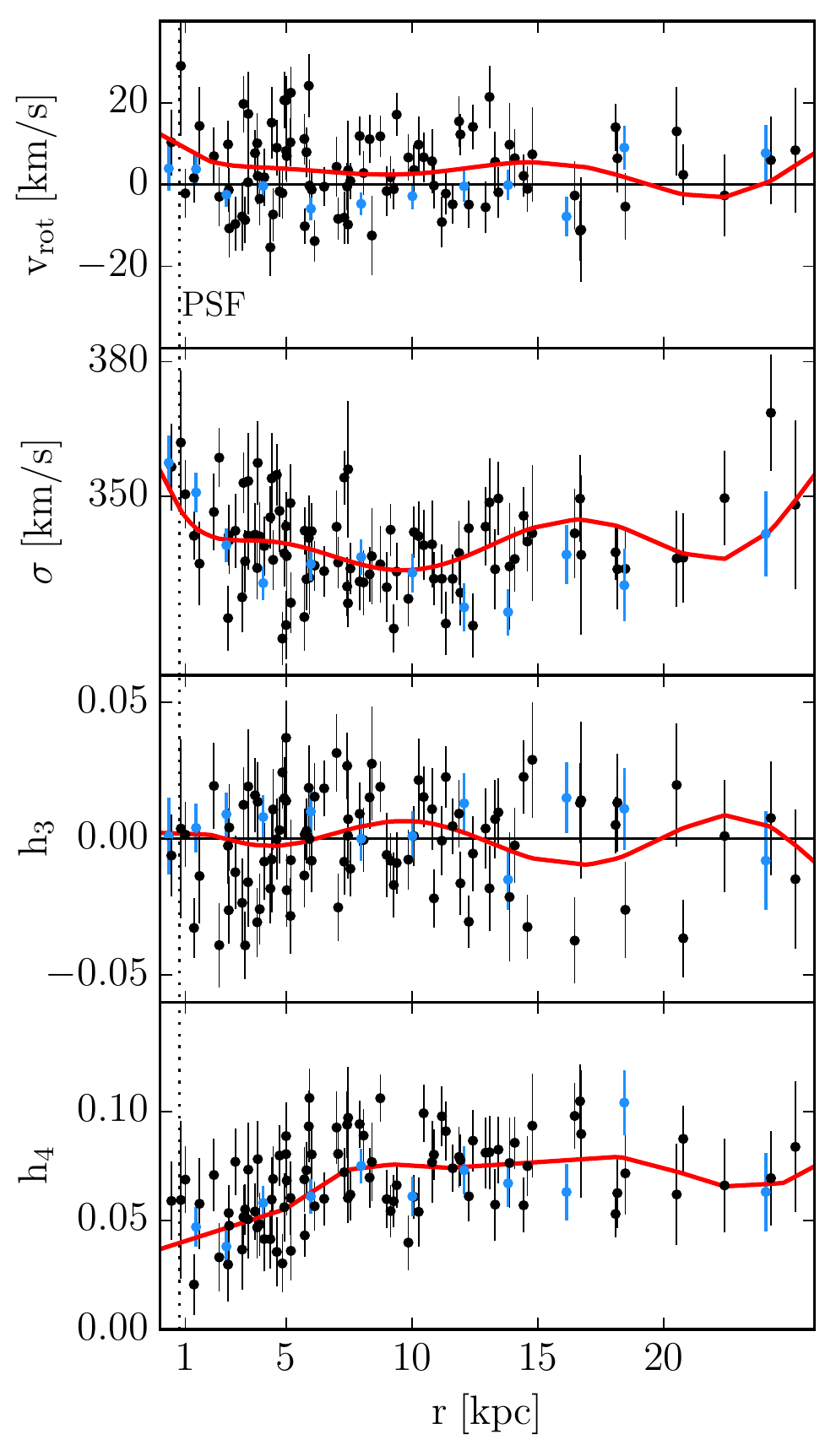}

	\centering
        \caption{The Gauss-Hermite moments measured from the MUSE spectra
(black points) compared to the best-fitting model (red). Shown
are (from top to bottom) $v_{\rm rot}$, $\sigma$, $h_{3}$, and $h_{4}$ of quadrant q4.
Note that the model was fit to the full non-parametric LOSVDs. The Gauss-Hermite moments are only used for illustrative purposes here.
The figure also includes Gauss Hermite moments measured independently on VIRUS-W spectra using the FCQ method (blue). 
}
	\label{fig:kinfit}
\end{figure}

The most important result from our dynamical modeling is the
detection of a SMBH with
$M_{BH} = (4.0 \pm 0.80) \times 10^{10} \si{M_{\odot}}$ in Holm~15A. The
associated SOI of this SMBH is
$r_{SOI} = 3.8 \pm \SI{0.37}{kpc}$ ($3\farcs5 \pm 0\farcs34$). Even though the galaxy is more than 200 Mpc away,
we spatially resolve the SOI by a factor of 10.  In
fact, $\sim 100$ out of our 421 LOSVDs sample the SOI of the galaxy. 
The modeling results for the black hole, stellar mass-to-light ratio and DM halo
parameters are summarised in Table ~\ref{tab:results}. $\Delta \chi^2$
curves for $M_{BH}$, $\Upsilon_{\star}$ and $\rho_{10}$ from all four
quadrants are shown in Figure \ref{fig:chi}. The figure shows that none of the four quadrants stands out and yields
a significantly different result than the others. While the black hole
mass in q3 (where the gas emission in the spectra is most prominent)
is slightly larger than in the other quadrants, this offset is not
significant. By computing the dynamical quantities separately for each
quadrant and estimating the uncertainties from these four nearly
independent measurements, we implicitly include any residual
systematics (like, e.g., from the gas emission) in our error budget.
Fits to the kinematics of one quadrant of
Holm~15A parameterized by $v_{rot}$, $\sigma$, $h3$ and $h4$ of our best-fit model are shown in
Figure \ref{fig:kinfit}. They show that our best-fit model can successfully reproduce the observed kinematics of 
the galaxy. For the non-parametric kinematics our best-fit model reaches a reduced $\chi^2$ of $0.8-0.9$ for each quadrant.

We had previously also acquired spectroscopy of Holm~15A from the McDonald Observatory using the low-resolution 
mode ($\sigma \sim \SI{25}{km/s}$) of the integral field unit  spectrograph VIRUS-W \citep{VIRUSW}.
Stellar kinematics for these independent data 
were derived by applying the Fourier Correlation method (FCQ) by \citet{Bender1990} 
in the wavelength interval between 4500 and 6250 \angstrom{}, using a sparser spatial sampling (Figure \ref{fig:kinfit}, blue) and circular spatial binning. This entirely independent measurement of 
the stellar kinematics in Holm~15A is consistent with the MUSE kinematics. We note that on average values of $h_4$ and $\sigma$ appear to be slightly lower for FCQ (likely due to a different smoothing-method). 
Therefore, as a consistency check we ran a second set of dynamical models using only the VIRUS-W kinematics and found the same results within the errors. Because the MUSE data have better spatial resolution and higher signal-to-noise ratio we will only discuss the results derived from the MUSE data in the remainder of this paper.

Finally, an example comparison between an observed and modelled LOSVD and a discussion
of the importance of the LOSVD wings can be found in Appendix \ref{sec:escdyn}.
\begin{table}
\centering
\begin{tabular}{@{}llr@{}}
Schwarzschild Model Parameter & Best-Fit Value  & Units\\ \toprule

$M_{BH}$  & $(4.0 \pm 0.80)$ & $10^{10} \ \si{M_{\odot}}$\\ 
$\Upsilon_{\star}$ (i-band) & $4.5 \pm 0.19$ & \\\hline
DM Halo: \\ 
$\rho_{10}$ &  $(1.0 \pm 0.10) $ & $10^{7} \ \frac{M_{\odot}}{kpc^3}$\\
$\log{r_{s}}$ &  $(2.4 \pm 0.29)$ & $\log{\frac{r}{kpc}}$\\
$\gamma$ & $0.35 \pm 0.26$ & \\
\end{tabular}
\caption{Results of Schwarzschild dynamical modeling of Holm~15A. 
  Best-fit values were derived as the mean of the independent fits 
    to the four quadrants. The quoted uncertainties 
    are derived from the variation between quadrants.}
\label{tab:results}
\end{table}

\section{Discussion}
\label{sec:discussion}
With $M_{BH} = (4.0 \pm 0.80) \times 10^{10} \si{M_{\odot}}$, the SMBH
at the center of Holm~15A is the most massive dynamically
determined black hole so far.  It is a factor of two larger than the
SMBHs in NGC~4889 \citep{2012ApJ...756..179M}, with
$M_{BH} = (2.1 \pm 0.99) \times 10^{10} \si{M_{\odot}}$ and NGC~1600 with
$M_{BH} = (1.7 \pm 0.15) \times 10^{10} \si{M_{\odot}}$
\citep{2016Natur.532..340T}. 
 Quasar luminosities at higher redshifts and current determinations of local SMBH scaling relations give an expected black hole cumulative space density
ranging from half a dozen up to a few hundred SMBHs with $M_{BH} \gtrsim 10^{10} \si{M_{\odot}}$ out to $z \leq 0.055$ 
\citep[e.g.][]{2007ApJ...662..808L, 2013AJ....146...45R}.
Hence, circumstances for the formation of a 40-billion-solar-mass SMBH
are probably rare, but the central structure of the Coma cluster
serves as an example that they do exist. As stated above, NGC4889, one of the two
central galaxies of Coma,  contains a SMBH of 
$M_{BH} = 2.1 \times 10^{10} \si{M_{\odot}}$. The other galaxy, NGC4874, has a very
extended classical shallow-power-law surface-brightness core with a
size of $r_b = 1.7 \, \mathrm{kpc}$ \citep{2007ApJ...664..226L}.  This
suggests a SMBH with a mass of
$M_\mathrm{BH} \sim 2 \times 10^{10} \, M_\odot$ (using the core
scaling relations of \citealt{2016Natur.532..340T}). Both galaxies are
in interaction and will eventually merge
 \citep[e.g.][]{2006IAUS..234..337A, 2007A&A...468..815G}. This will produce a BCG at
the center of the Coma cluster which will very likely have a SMBH 
in the same mass range as Holm~15A has now.

In the following sections we will discuss the observational and theoretical evidence for the merger origin of Holm~15A, as well as attempt to unravel some specific details of the merger history.

\subsection{SMBH-scaling relations: Evidence for dissipationless merging}
\label{sec:scaling}

\begin{figure*}
      \includegraphics[width=2\columnwidth]{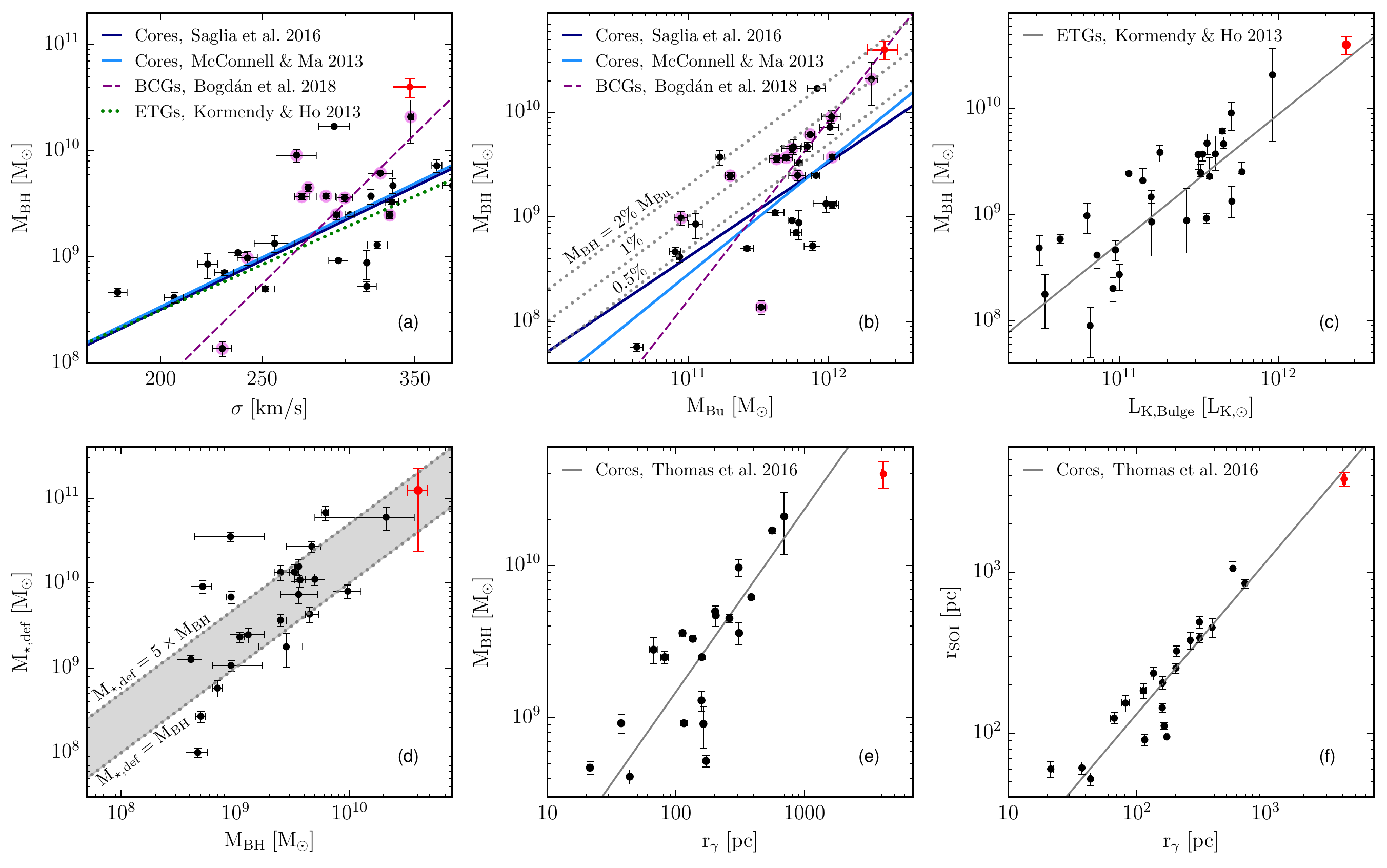}
	\centering
       \caption{Holm15A (red) compared to other ETGs, Cores and BCGs (black) on SMBH-scaling relations.
       \textbf{a \& b}: Holm~15A compared to cored ETGs listed in \citet{2016ApJ...818...47S}  with respect 
       to the global galaxy scaling relations, $M_{BH} - \sigma$ (\textbf{a}) and $M_{BH} - M_{Bu}$ (\textbf{b}).
	  Solid lines show the linear relations for 
	  cored ETGs from 
	   \citet{2016ApJ...818...47S} and 
	  \citet{mcConnell&ma2013}. Dashed and dotted lines indicate scaling relations for ETGs in general (cored or not) from \citet{2013ARA&A..51..511K} and 
	  BCGs-only from \citet{Bogdan2018}. ETGs identified as BCGs in \citet{Bogdan2018} are enhanced by purple halos around their symbols.
	  \textbf{c}: Holm~15A's directly measured (3KK) K-band luminosity $L_K$ compared to ETGs from  \citet{2013ARA&A..51..511K} on the global galaxy scaling relation $M_{BH}-L_K$. 
	   The line shows the linear relation from \citet{2013ARA&A..51..511K}.
	  \textbf{d}: Core-mass deficits $M_{\star, def}$ of cored ETGs from \citet{2013AJ....146..160R} and Holm~15A. 
	  \textbf{e \& f}: Holm~15A compared to cored ETGs from  \citet{2016Natur.532..340T} and
       \citet{2013AJ....146...45R, 2013AJ....146..160R} with respect to
        the 
	  core-specific scaling relations, $M_{BH} - r_{\gamma}$ (\textbf{e}) and $r_{SOI} - r_{\gamma}$ (\textbf{f}).
	   The lines show the linear relations from \citet{2016Natur.532..340T}. The figure includes 
	   the uncertainties of $r_{\gamma}$, but they are generally smaller than the symbol size.
	  }      
	\label{fig:ScalingSummary}
\end{figure*}

The SMBH of Holm~15A is not only the most massive one to date, it is
also four to nine times larger than expected given the galaxy's stellar mass
$M_{Bu} = (2.5 \pm 0.64)  \times 10^{12} \si{M_{\odot}}$
and the
galaxy's stellar velocity dispersion
$\sigma = (346 \pm 12.5) \ km/s$
(see Figure \ref{fig:ScalingSummary}a).

It has been previously noted that the $M_{BH} - \sigma$ relation may shallow out at the high-mass
end due to dry merging becoming the dominant growth process at the high-mass end. Since dry (major) mergers grow $\sigma$ only slowly \citep[e.g.][]{2007ApJ...662..808L, Naab2009}
but simply sum over the central SMBH masses of the merging galaxies, such mergers will move galaxies towards ``overmassive'' $M_{BH}$ at a given $\sigma$ \citep[e.g.][]{2007ApJ...662..808L, 2013ApJ...769L...5K}. Correspondingly, massive core galaxies 
follow a $M_{BH} - \sigma$ relation that is steeper and slightly offset (towards larger values of $M_{BH}$) compared to less massive, cuspy galaxies (cf.  \citealt{2016ApJ...818...47S} and \citealt{mcConnell&ma2013}).
Despite the fact that we here already consider the $M_\mathrm{BH}-\sigma$ relation of core galaxies,
 Holm~15A is still almost an order of magnitude offset in $M_{BH}$  (see Figure \ref{fig:ScalingSummary}a). This might be indicative
of an especially extensive dry merging period.

One could expect the
$M_{BH} - M_{Bu}$ relation to be tighter at the high-mass
end, since the ratio $M_\mathrm{BH}/M_\mathrm{Bu}$ ratio is conserved in dry mergers. Holm~15A, however, is also a strong outlier from this relation ($M_{BH}$ is roughly 4 times larger than expected from $M_{Bu}$, see Figure \ref{fig:ScalingSummary}b). The ratio between $M_{BH}$ and $M_{Bu}$ is typically $\lesssim 0.5 \%$ for
cored ETG and typically $\lesssim 1 \% $ when considering all ETGs below a stellar mass of  $< 10^{13} M_{\odot}$, \textit{irrespective} of central morphology  \citep{2013ARA&A..51..511K}. Holm~15A, however, hosts a black hole that contains close to $2 \%$ of the total stellar mass of the galaxy. A similar high ratio as been found in NGC~1600 \citep{2016Natur.532..340T}.
This might suggest that the progenitor galaxies of Holm~15A were different from typical massive ETGs at $z \sim 0$. Studies of the evolution of $M_\mathrm{BH}/M_\mathrm{Bu}$ since $z \sim 3$ in active galaxies suggest that the ratio scales like $(1+z)^{0.7-1.4}$ \citep[e.g.][]{Decarli2010, Merloni2010, Bennert2011}. Depending on which 
$M_{BH} - M_{Bu}$ relation is used (all central morphologies or cores-only) we can estimate 
that Holm15A's progenitors might have formed early, at $z \gtrsim 1$ or $2$. 

In Fig.~\ref{fig:ScalingSummary}b we only consider scaling relations based on dynamcial bulge masses to avoid systematics related to assumptions about the initial stellar mass function (IMF). We will touch on this again in Sec.~\ref{sec:DM}. 

\citet{Kluge2019} showed that BCGs and ETGs in general follow different scaling relations between total luminosity, size and effective surface brightness. This would also translate into different SMBH scaling relations. 
 \citet{Bogdan2018} suggested that BCGs follow steeper $M_{BH}-\sigma$ and $M_{BH}-M_{Bu}$  
 relations (cf. 
Fig.~\ref{fig:ScalingSummary}a,b). Holm~15A is closer to these BCG-centric scaling relations. In fact, it happens to fall onto the corresponding $M_{BH}-M_{Bu}$ relations and is offset from the corresponding $M_{BH}-\sigma$ relations by about a factor of two. This could indicate that the galaxy formed from a dissipationless, (roughly) equal-mass BCG-merger, though the scatter in the relations is large. 

 We note that the total stellar mass of Holm~15A is estimated based on the assumption that the mass-to-light ratio is constant out to a region that is almost 10 times larger than the field of view of our kinematic observations. Therefore, in Fig.~\ref{fig:ScalingSummary}c, we also compare Holm~15A's K-band luminosity $L_{K}$ to the $M_{BH}-L_{K}$ relation of \citet{2013ARA&A..51..511K}. $L_K$ was measured from an image that extends out to $\sim \SI{250}{kpc}$ and that was obtained with the three-channel imager at the Wendelstein 2-m Telescope \citep[3KK][]{3KK2010, 3KK2016}.
Holm~15A follows the $M_{BH}-L_{K}$ correlation better than the $M_{BH}-M_{Bu}$ relation.

\begin{figure}
      \includegraphics[width=1\columnwidth]{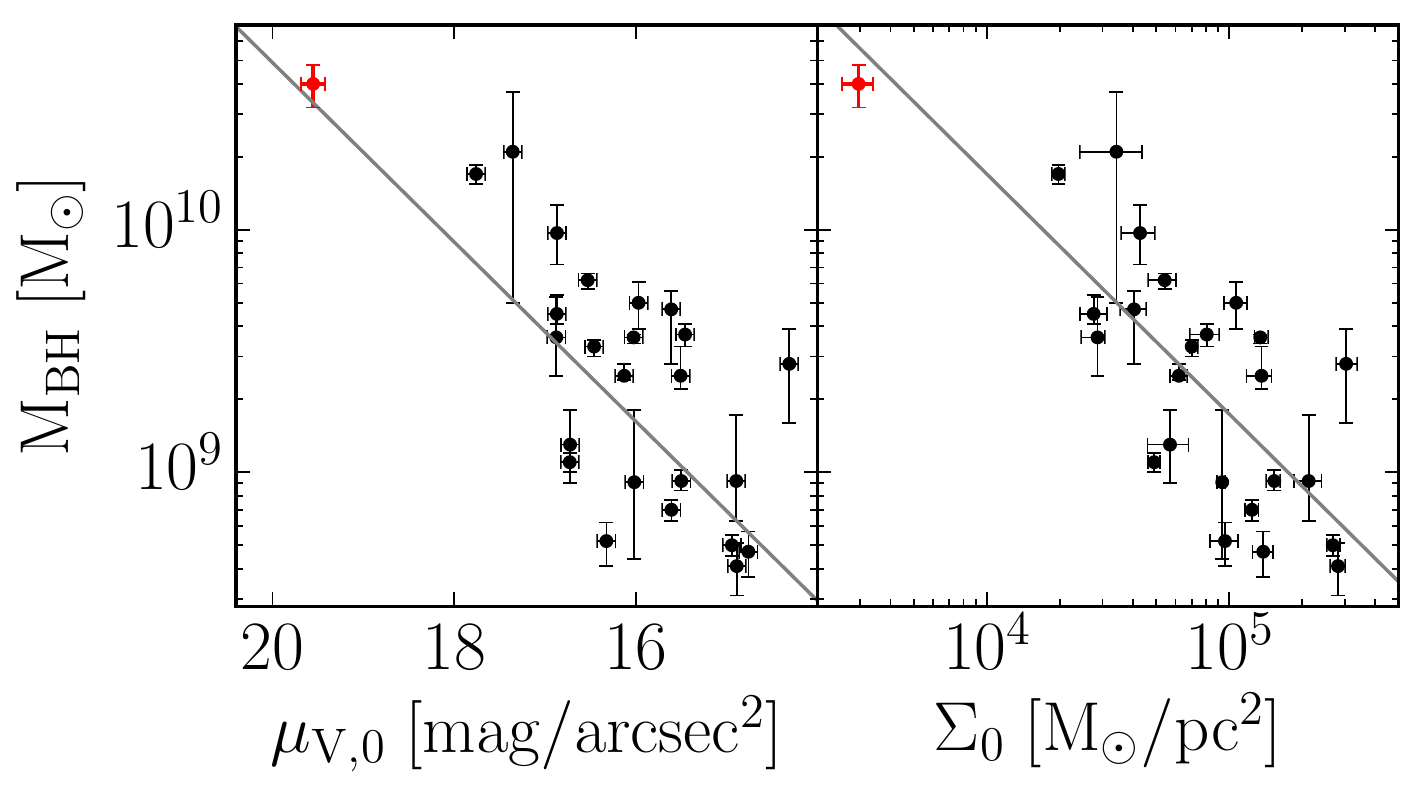}
	\centering
       \caption{The central V-band surface brightness $\mu_{V,0}$ (left) and 
       stellar surface mass density $\Sigma_0$ (right) versus $M_{BH}$ for Holm~15A (red) and cored ETGs from \citet{2013AJ....146..160R, 2013AJ....146...45R} and NGC1600 \citep{2016Natur.532..340T} (black).
        All values of $\mu_{V,0}$ and $\Sigma_0$ relate directly to the \textit{observed} light profiles themselves.
	    The line shows the best-fit linear relation.
		}
	\label{fig:surfdens}
\end{figure}

\subsection{Scaling relations of core properties: similarity with other core galaxies}
\label{sec:corescaling}
Dissipationless mergers between ETGs involve binary black hole core-scouring and, hence, result in depleted,
low-surface-brightness cores. 
As already mentioned above, core galaxies follow specific scaling relations
between the core size, missing light and black-hole mass \citep{2007ApJ...664..226L, 2013AJ....146..160R} 
and the radius
of the sphere-of-influence \citep{2016Natur.532..340T}. Simulations have shown that these relations can be explained by the black-hole binary model \citep{2018ApJ...864..113R}. 
In Figure \ref{fig:ScalingSummary}d we show the central stellar mass deficits from \citet{2013AJ....146..160R} together with Holm~15A. The mass deficit in Holm~15A 
is $M_{\star, def} = (1.24 \pm 1.00) \times 10^{11} M_{\odot}$, based on the dynamical stellar mass-to-light ratio
and $L_{i, def}$  derived in Sec.~\ref{sec:missinglight} . This roughly corresponds to $0.5 - 5.5$ times the black hole mass, similar to the mass deficits in many other core galaxies.

In Figure \ref{fig:ScalingSummary}e we compare the core size of Holm~15A to other galaxies. As described in Sec.~\ref{sec:missinglight} we use the cusp radius $r_{\gamma}$ here. Compared to the 
galaxies of \citet{2013AJ....146...45R, 2013AJ....146..160R, 2016Natur.532..340T} the core in
Holm~15A is roughly a factor $2.5$ larger than expected for the mass of its black hole.

Such an offset could be explained, for example, if Holm~15A experienced an early phase of rapid 
evolution with an enhanced merger rate. It could well be then that not only a binary
black hole was involved in the formation of its core, but possibly a more complicated system of
multiple black holes.  Theory suggests that core scouring efficiency
is significantly enhanced by multiple black holes and that cores grow
much larger \citep{2012MNRAS.422.1306K}.  We will revisit this issue in Sec.~\ref{sec:nbody}. In Figure \ref{fig:ScalingSummary}f we compare $r_{\gamma}$ with the radius of 
the sphere of influence $r_{SOI}$. Despite being offset on the $M_{BH}-r_{\gamma}$ relation the cusp radius is 
consistent with the correlations between core-size measurements and $r_{SOI}$ 
in other core galaxies

\subsection{A new correlation between black-hole mass and core surface brightness}
\label{sec:newscaling}
Cores in massive ETGs obey a strong homology in that the central
surface brightness correlates inversely with
the size of the core \citep{1997AJ....114.1771F, 2007ApJ...662..808L} -- This, together with the scaling between $M_{BH}$ and core size, implies a
potential
scaling between $M_{BH}$ and the central
surface brightness $\mu_{0}$ in cores. An equivalent argument can be made for a 
correlation between $M_{BH}$ and central stellar surface mass density $\Sigma_{0}$.
We show these correlation in Figure~\ref{fig:surfdens} for the galaxy sample of \cite{2013AJ....146...45R}, NGC1600 \citep{2016Natur.532..340T} and Holm~15A.
We used the uncertainties for the stellar mass-to-light ratios and black hole masses listed in 
\citet{2013AJ....146..160R,2013AJ....146...45R} and \citep{2016Natur.532..340T} and assumed rather conservative uncertainties
of $\SI{0.1}{mag/arcsec^2}$ for the light profiles.
Our best-fit linear relations were determined following the approach to linear regression from  \citet{Kelly2007} (using the Python package \textit{linmix} by \citealt{linmix}) with errors in both $M_{BH}$ and $\mu_{V,0}, \Sigma_{0}$:
\begin{equation}
\begin{aligned}
 \log{(M_{BH}/\si{M_{\odot}})} ={} & (0.37 \pm 0.07)  \mu_{V, 0} \si{mag^{-1} arcsec^{2}} \\
                                   & + (3.29 \pm 0.37)
 \end{aligned}
\end{equation}

 \begin{equation}
 \begin{aligned}
 \log{(M_{BH}/\si{M_{\odot}})} ={} & (- 0.99 \pm 0.19)\log{(\Sigma_{0}/\si{M_{\odot} pc^{-2}})} \\
                                   &+ (14.19 \pm 0.09).
  \end{aligned}
 \end{equation}
The $M_{BH}-\mu_{V, 0}$ relation has an intrinsic scatter $\epsilon = 0.32 \pm 0.07$. Similarly,
the $M_{BH}-\Sigma_{0}$ relation has an intrinsic scatter of $0.30 \pm 0.07$.
Values of $\Sigma_{0}$ were 
calculated from the surface brightness at the spatial resolution limit for each galaxy and
their corresponding dynamical stellar mass-to-light ratios \citep{2013AJ....146..160R,2013AJ....146...45R, 2016Natur.532..340T}. Values for both $\mu_{V, 0}$ and $\Sigma_{0}$ were determined using the \textit{observed} light profiles of each core galaxy.
Holm~15A has the lowest central stellar surface brightness/mass, $\mu_{V,0}$ = $19.9 \pm \SI{0.13}{mag/arcsec^{2}}$, $\Sigma_{0} = (3.0 \pm 0.40) \times 10^3 \si{M_{\odot}/pc^{2}}$ of all core galaxies with dynamical black hole mass measurements (cf. Figure \ref{fig:surfdens}).
Nonetheless, Holm~15A is fully consistent with the homology established by other core galaxies. \footnote{The listed relations were determined \textit{including} Holm~15A, but the relations change only marginally and within the listed uncertainties 
when we exclude the galaxy.}
All of the above evidence points to the fact that the core in Holm~15A was formed by the
same physical process as cores in other massive ETGs, i.e. by a black-hole binary.

\subsection{N-body merger simulations: evidence for a merger between two core galaxies}
\label{sec:nbody}
We will now discuss what the specific photometric and orbit-dynamical properties of Holm~15A may tell us about its merger history. 

\begin{figure}
      \includegraphics[width=0.8\columnwidth]{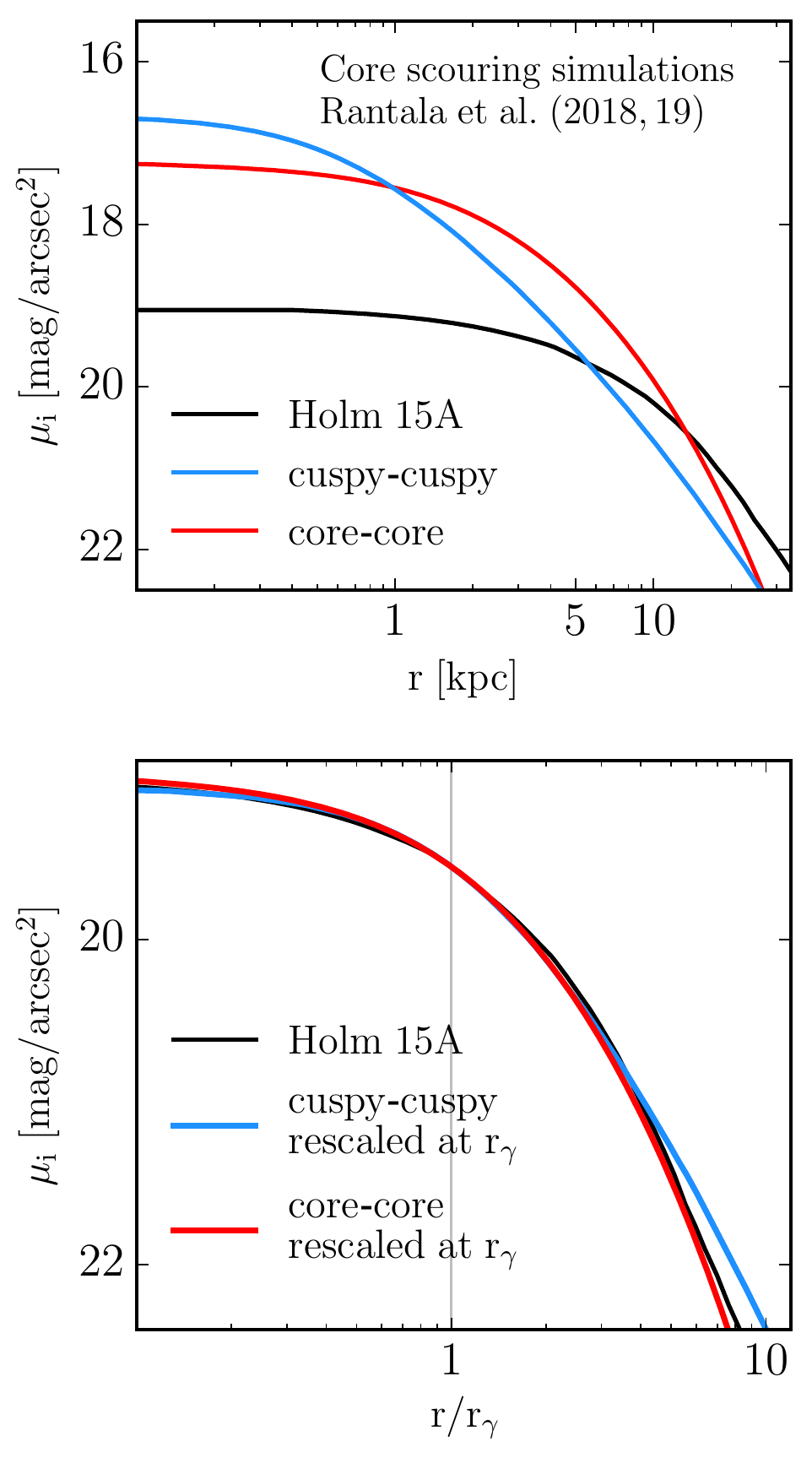}
	\centering
        \caption{Top panel: $i$-band surface brightness profile of
          $\mu (r)$ of Holm~15A (black) compared to the remnants of
          numerical merger simulations with core scouring. The blue
          profile shows a merger between two cuspy galaxies with a
          final black hole mass of
          $M_{BH} = 1.7 \times 10^{10} \ \si{M_{\odot}}$, roughly half
          of the black-hole mass observed in Holm~15A.  The red profile
          is the result of remerging this remnant with itself,
          doubling the mass of the central black hole to
          $M_{BH} = 3.4 \times 10^{10} \ \si{M_{\odot}}$. Bottom
          panel: Holm~15A compared to the remnant surface brightness
          profiles scaled to the value $\mu (r \equiv r_{\gamma})$ of
          Holm~15A.}
	\label{fig:CoreCoreSB}
\end{figure}

In Figure~\ref{fig:CoreCoreSB} we
compare the light profile of Holm~15A with the N-body merger simulations of
\citet{2018ApJ...864..113R,2019ApJ...872L..17R}.  These simulations study the outcome of
a dissipationless merger between two early-type progenitor galaxies,
both with central black holes. The simulations follow the dynamical
interaction between the black hole binary that temporarily forms at
the center of the remnant galaxy and the surrounding stars with
high accuracy.  The figure demonstrates that mergers between cuspy
progenitors (i.e. mergers between originally coreless progenitor
galaxies) lead to slightly different light profiles than do mergers
between galaxies that already had cores.  The light profile of Holm
15A, in fact, looks very similar to the 2nd type of merger,
i.e. between two already cored galaxies\footnote{At roughly
  $8 \times r_{\gamma} \sim \SI{40}{kpc}$ (for Holm~15A) the surface
  brightness of the rescaled core-core remnant drops faster than the
  that Holm~15A.  This could be due to the fact that the merger
  simulations do not include an extended cD halo. Photometric studies
  of Holm~15A \citep[e.g.][]{Kluge2019, 2011ApJS..195...15D}
  suggest an extended stellar envelope starting at
  $r \gtrsim \SI{35}{kpc}$.  At radii < $8 \times r_{\gamma}$ the
  core-core remnant is remarkably similar to Holm~15A.}
(Figure~\ref{fig:CoreCoreSB}).
        
\begin{figure}
      \includegraphics[width=0.9\columnwidth]{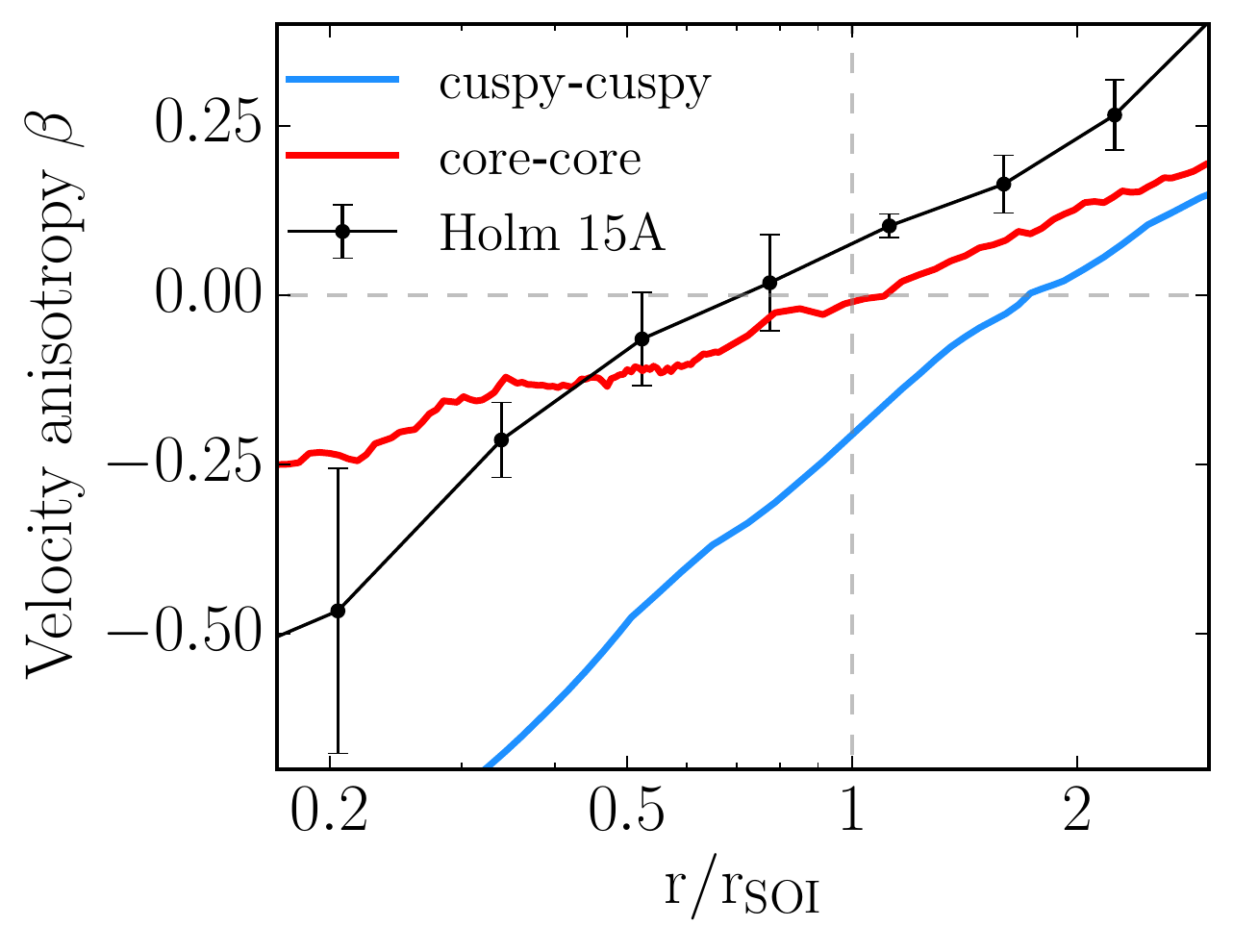}
	\centering
        \caption{Anisotropy profile $\beta (r)$ of our best-fit
          dynamical model of Holm~15A averaged over four quadrants
          compared to numerical merger simulations of binary black hole
          core scouring
          from Figure \ref{fig:CoreCoreSB}, in the same colors as before.
          Radii are scaled by $r_{SOI}$.}
	\label{fig:ani}
\end{figure}

The evidence in favor of a core-core merger from the light profile is consistent with
the evidence from the orbit distribution that we find in
Holm~15A. Figure~\ref{fig:ani} shows the radial profile of the anisotropy
parameter
\begin{equation}
\beta  = 1 - \frac{\sigma_{t}^2}{\sigma_{r}^2},
\label{eq:beta}
\centering
\end{equation}
where $\sigma_{r}$ is the radial and
$\sigma_{t} = \sqrt{(\sigma_{\theta}^2 + \sigma_{\phi}^2)/2}$ is the
tangential velocity dispersion, computed from the dispersions
$\sigma_{\theta}$ and $\sigma_{\phi}$ in the two angular
directions. The figure also includes the results from the
numerical N-body simulations.  It is known that core scouring results
in an orbital distribution that is biased increasingly towards
tangential orbits ($\beta < 0$) inside the SOI of the black hole as $r \rightarrow 0$
and increasingly towards radial orbits ($\beta > 0$) outside of it, towards
larger radii \citep[e.g.][]{1997NewA....2..533Q,
  2001ApJ...563...34M, 2018ApJ...864..113R}.
  Tangential anisotropy around SMBHs has been observed in systems
  of various masses and morphologies \citep[e.g.][]{verolme02,gebhardt03,shapiro06,houghton06,gebhardt09,gueltekin09,krajnovic09,siopis09,shen10,vandenbosch10,schulze11,gebhardt11,mcconnell12,walsh15,feldmeier17,2016Natur.532..340T}. In core
  galaxies, specifically, the measured anisotropy is extremely homogeneous and intimately linked
  to the core region and follows very closely the prediction of N-body merger simulations \citep{2014ApJ...782...39T}.
  
  In Holm~15A we see the same behaviour: a change from outer radial anisotropy to inner tangential motions
  roughly at the sphere of influence radius (which is similar to the core size, see Figure \ref{fig:ScalingSummary}f). The evidence for this comes from the wings of the observed LOSVDs (cf. App.~\ref{sec:escdyn}).
  However, the central anisotropy in Holm~15A is milder than observed in other core galaxies, which follow the ``cuspy-cuspy'' line in Figure~\ref{fig:ani} \citep{2018ApJ...864..113R}. This difference is actually expected if the direct progenitors
  of Holm~15A were not cuspy power-law ellipticals but galaxies that already had cores. In the latter case, the anisotropy
  in the center is predicted to be very similar to the observed orbital structure of Holm~15A \citep{2019ApJ...872L..17R}\footnote{In the N-body simulations, the final
anisotropy profile of an equal-mass core-core merger is very similar
to that of the final orbit distribution after a sequence of minor mergers
\citep{2019ApJ...872L..17R}.  However, the light profile of Holm~15A is
more similar to the core-core merger than to the remnant after
repeated minor mergers. Further simulations covering a wider range of
initial conditions are needed to confirm the connection between
anisotropy, profile shape and merger history.}.

Since cores grow with each merger generation, a core-core merger scenario
would plausibly explain the fact that the central region of
Holm~15A is fainter than the centers
of $\gtrsim 97\%$ of the 164 local ETGs in
\citet{2007ApJ...664..226L}, despite the fact that the galaxy is more
luminous than $\gtrsim 90\%$ of the sample ($M_{V} = -23.8 \pm 0.1$,
\citealt{2014ApJ...795L..31L}; see also Figure~\ref{fig:surfdens}). It would also explain the large core
size of Holm~15A. 

Moreover, it could even provide a reason for Holm~15A's
large cusp radius (Figure~\ref{fig:ScalingSummary}e): In the merger simulation
during the core-core re-merger $M_{BH}$ doubled while the core radius (described either by $r_{b}$ or $r_{\gamma}$) roughly tripled in size. This would suggest that in successive core scouring events the core grows faster then the central black hole. Similarly, for a sequence of five smaller core scourings due to minor mergers, the remnant also ``outgrew'' its black hole by a similar factor.

In the merger case, Holm~15A
represents a dynamically very evolved galaxy that is possibly one
merger generation ahead of cored galaxies like NGC4874 and NGC4889 at the
center of the Coma cluster. 
As we showed in the previous subsection, Holm~15A's
high $M_{BH}/M_{Bu}$ ratio of $\sim 2\%$ might indicate that
the galaxy's progenitors had already formed at redshifts larger than 1 or 2 and/or that its progenitors were themselves BCGs.
Abell~85 has one of the strongest cool-cores among X-ray bright clusters \citep{Chen2007} and is strongly BCG dominated, with
Bautz-Morgan morphological type I \citep{2010A&A...513A..37H} such
that the central parts of the main cluster in fact might have been 
subject to a slightly accelerated evolution at some point in the past.
Previous X-ray studies of Abell 85 had already 
suggested that the measured temperature and metallicity maps of the cluster
were compatible with an intense merger history \citep[e.g.][]{2005A&A...432..809D, 2010ApJ...721.1262M}.

\subsection{Alternative formation scenario via AGN feedback?}
\label{sec:holmagn}
Even though the merger scenario provides a consistent
explanation for the central light profile shape of the galaxy, 
the orbital structure and how both are connected to the mass of
the central black hole, we briefly discuss whether the interaction between an AGN and
the surrounding stars could serve as an alternative core-formation scenario. 

In recent simulations, AGN outflows have been
observed to trigger fluctuations of the local gravitational potential
which irreversibly transfer energy to the dark matter and stellar
components \citep{2011MNRAS.414..195T, 2012MNRAS.422.3081M,2013MNRAS.432.1947M, Choi+2018}. 
These simulations produced exponential light
profiles, which resemble the cores of ETGs in the sense
discussed in the introduction: the central surface brightness is low
and the slope of the central surface brightness profile is
shallow.  In fact, based on the black-hole
fundamental plane it has been argued that many black holes in the BCGs of
cool-core clusters could be more massive than predicted by the
classical black-hole scaling relations, and many would actually be
expected to have masses $M_\mathrm{BH} > 10^{10} \, M_\odot$ \citep{2019ApJ...875..141P, 2012MNRAS.424..224H}.  We are still lacking numerical
simulations that study in quantitative detail the effect of AGN
feedback on the stellar light distribution and orbital structure.
The information contained in the actual orbits
of the stars might turn out to be crucial to distinguish between
different core formation scenarios.
\subsection{Dark matter halo and stellar mass-to-light ratio}
\label{sec:DM}
\begin{figure}
      \includegraphics[width=1\columnwidth]{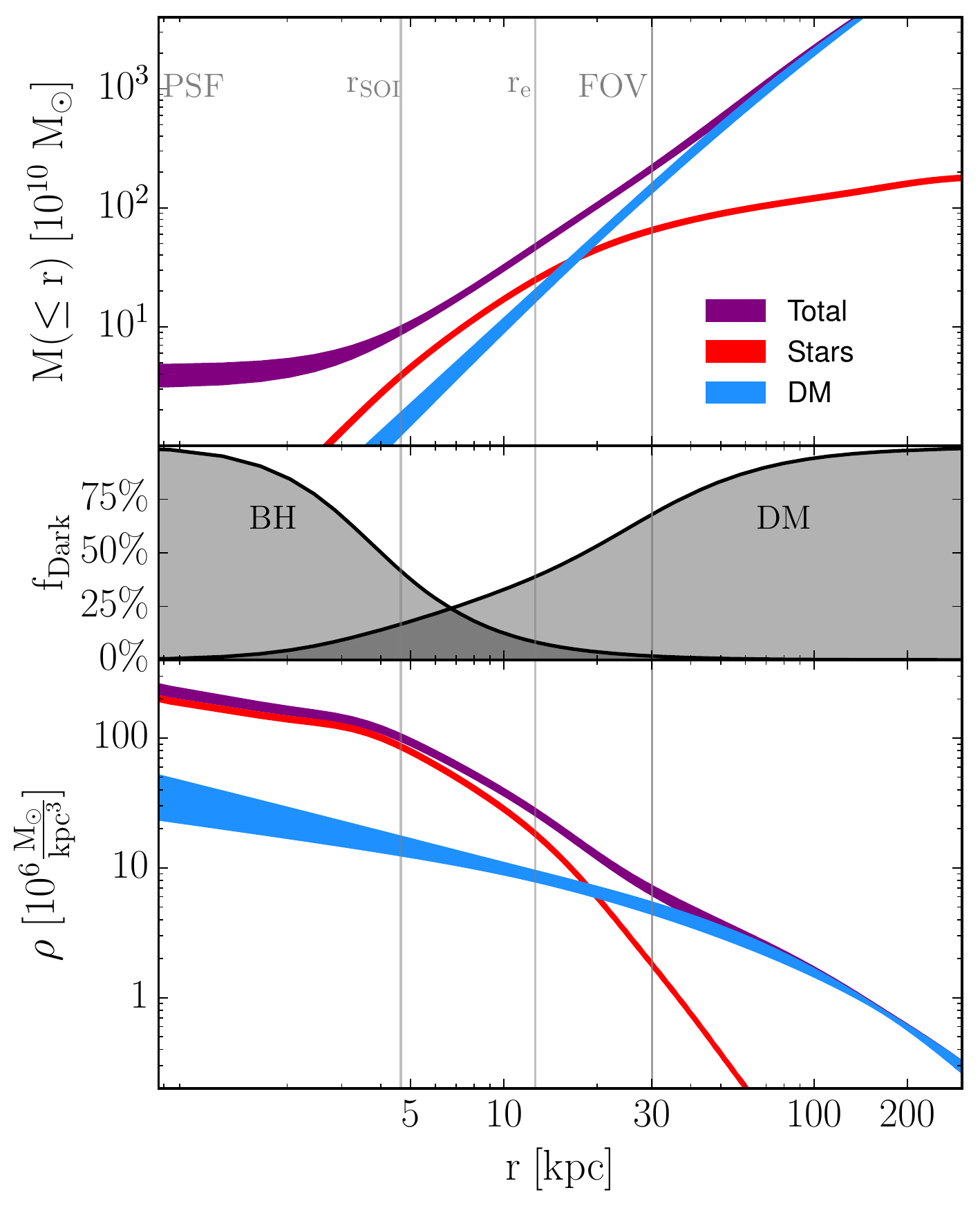}
	\centering
       \caption{Top panel: enclosed mass profile of the best-fit dynamical model of Holm~15A,
       separated into total (including black hole, purple), stellar (red) and DM (blue) mass. The broadness of the profiles
       indicates the variation of best-fit models between the quadrants.
       The middle panel indicates the fraction of non-luminous 
       mass, i.e. the black hole and DM halo, with respect to the total enclosed mass at a 
       given radius for the best-fit model. 
       The bottom panel shows corresponding the stellar, DM and total density distributions. 
       }
	\label{fig:BestFitDynamics}
\end{figure}

Figure \ref{fig:BestFitDynamics} shows the underlying stellar, dark
matter and total enclosed mass and density profiles of our best-fit
dynamical model of Holm~15A. Apart from the $20\%$ variation in $M_{BH}$, the quadrants 
 of the galaxy produce a consistent overall mass and density profile.  

Using simple stellar population models \citep{2003MNRAS.339..897T, Maraston+11} we find that Holm~15A has a marginally super-solar metallicity, $[Z/H] = 0.08 \pm 0.05$ and is strongly $\alpha$-enhanced $[\alpha/Fe] = 0.25 \pm 0.03$. Assuming a Kroupa stellar
initial mass function (IMF) we find a stellar mass-to-light ratio 
of $\Upsilon_{SSP, Kroupa} = 2.7 \pm 0.30$ (i-band) using methods from either
\citet{Maraston+11} or \citet{2017ApJ...837..166C}.
The large $\sim 20 \%$ uncertainty of this value is due to the difficulty of determining the age of the stars. Formally, our 
SSP models fitted stellar ages that exceed the age of the universe. The value of $\Upsilon_{SSP, Kroupa}$ and its uncertainty are derived from ``manually'' varying stellar ages between $\SI{10}{Gyrs}$ and $\SI{13.8}{Gyrs}$ while fixing elemental abundances.

Our dynamical mass-to-light ratio of $\Upsilon_{\star} = 4.5 \pm 0.19$ is
roughly twice as large as the SSP ratio 
($\Upsilon_{\star} / \Upsilon_{SSP,Kroupa} = 1.7 \pm 0.20$).
This is a continuation of a growing trend among recent
mass-to-light ratio measurements in massive ETGs from dynamics,
lensing and spectroscopy often finding values larger than predicted by SSP
models adopting a Kroupa stellar IMF,
$\Upsilon_{\star} / \Upsilon_{SSP, Kroupa} \gtrsim 1.6$ \citep[e.g.][]{treu10,auger10,2011MNRAS.415..545T,
  2011MNRAS.417.3000S, 2012Natur.484..485C,
  2012ApJ...760...71C,tortora14,2017ApJ...837..166C,parikh18,alton18}. This offset is roughly consistent with a mass-to-light ratio implied by Salpeter-like IMF or might suggest that DM traces the stars. Our stellar-dynamical mass-to-light ratio is based on the assumption that all mass tracing the galaxy's light profile belongs to the stars of the galaxy. 
  In this case, when parameterizing the inner DM-halo as
$\rho_{DM} \sim r^{-\eta}$, we find $\eta = 0.45 \pm 0.16$ out to roughly
$\SI{50}{kpc}$.
This is substantially shallower than predicted by numerical simulations of cold dark
matter, $\eta \geq 1$ \citep[e.g][]{1996ApJ...462..563N,
  1997ApJ...490..493N, 1998ApJ...499L...5M}.  Combined stellar
kinematics and weak \& strong lensing studies of local BCGs previously found
$\rho_{DM} \sim r^{-0.5}$ on scales comparable to the effective radius
\citep[e.g][]{2004ApJ...604...88S,
  2008ApJ...674..711S,2013ApJ...765...25N}. 
  
  Within the core region the fraction of DM is $\lesssim 20 \%$. However, under the assumption of a Kroupa IMF and that DM traces stars, 
    the fraction of DM within the core region would be roughly $50 \%$, while in the former scenario equality between the enclosed stellar and DM mass is reached only
    at $r_{eq} = 33 \pm \SI{2.5}{kpc}$
    (The stellar mass density
profile reaches equality with the DM density profile at $28 \pm \SI{0.10}{kpc}$).  In both scenarios the mass density distribution of the stars in our best-fit model has a slope similar to that of
     the distribution of DM inside the core, $\rho_{total} \sim r^{-0.5}$.
  
 We note that some massive galaxies seem consistent with a low-mass IMF \citep[e.g.][]{2016Natur.532..340T,collier18} and that some fine-tuning is required to consistently
 combine masses from
multiple constraints like lensing, dynamics or spectroscopy  \citep[e.g.][]{2017ApJ...845..157N}. Dynamical and lensing constraints, in general, become model dependent when stars and DM trace each other closely  \citep[e.g.][]{2011MNRAS.415..545T}.

\section{Summary and Conclusions}
\label{sec:conclusion}
We have observed Holm~15A, the BCG of the cool-core galaxy cluster Abell~85, with
MUSE. Our observations reveal a galaxy with little rotation ($v_{rot} < 40 \, \mathrm{km/s}$) and a nearly constant velocity
dispersion of $\sigma = 340 \, \mathrm{km/s}$. Towards the center
and towards large radii, the velocity dispersion increases slightly.

We use orbit-based, axisymmetric Schwarzschild models to analyse the
dynamical structure of Holm~15A and compare them to recent
high-resolution N-body simulations of mergers between ETG galaxies that
host black holes.  Our results indicate the following:
\begin{itemize}
\item Holm~15A hosts a $(4.0 \pm 0.8) \times 10^{10} \ \si{M_{\odot}}$ SMBH
  at its center, the most massive black hole directly deteced via stellar
  dynamics so far. The black hole constitutes close to $2 \%$ of the
  total stellar mass of the galaxy.

\item Inside of the gravitational sphere of influence of the black hole,
  $r_{SOI} = 3.8 \pm \SI{0.37}{kpc}$, the orbital distribution becomes
  increasingly tangentially anisotropic. However, the anisotropy
inside the core is less tangential than in other big elliptical galaxies with depleted cores.
\item The galaxy's light profile and the observed mild orbital anisotropy both match remarkably well with predictions from N-body simulations of a merger between two elliptical galaxies that already had depleted cores.

\item The SMBH is roughly 9 times larger than expected from the  
      $M_{BH} - \sigma$ relation and 4 times larger than expected from the stellar mass
     of the galaxy, when compared to other cored ETGs. However, the offsets are smaller when compared to other BCGs.

\item In core galaxies black hole masses scale inversely with the central stellar surface brightness $\mu_{0}$ and central stellar mass density $\Sigma_{0}$ - including in Holm~15A. We show this correlation here for the first time.

\item Even in extreme instances of core formation like in Holm~15A, the core-specific relations $M_{BH} - \Sigma_{0}$, $M_{BH} - \mu_{0}$, $r_{SOI} - r_{\gamma}$, as well as the global galactic relation $M_{BH} - L_{K}$ still seem to hold. But the details of the light profile and orbital anisotropy contain valuable information about the specific formation path.

\item Assuming that all the mass that follows the light is stellar, we infer a bottom-heavy IMF,  $\Upsilon_{\ast} = 4.5 \pm 0.19$ ($i$-band), and the inner power-law slope of the DM-density distribution to be $\eta = 0.45 \pm 0.16$. Equality between enclosed stellar and DM mass is reached at $33 \pm \SI{2.5}{kpc}$. Assuming a Kroupa IMF, $\Upsilon_{SSP, Kroupa} = 2.7 \pm 0.3$, and DM tracing stars, we infer $\eta \sim 1$ outside of the core and a DM-fraction of nearly $50\%$ within the core.
\end{itemize}

We plan to extend our analysis of the galaxy to triaxial Schwarzschild models.
This will allow us to investigate potential systematics related to symmetry assumptions in the modelling and related to possible substructure near the very center of the galaxy. 

Our results suggest that the exact shape of the central light profile
as well as the details of the distribution of stellar orbits in the
center contain valuable information about the merging history of very
massive galaxies. E.g., extreme instances of core formation could
potentially lead to remnant surface-brightness profiles diverging from
the typical core-S\'{e}rsic profiles of ``classical'' cored galaxies.
Hydrodynamical
cosmological simulations have also produced large stellar and
dark-matter cores through AGN feedback. It will be interesting to
compare the anisotropy profiles predicted by these simulations with
measurements in observed galaxies.  More extensive simulations are
also required to investigate in detail the effect of core scouring
under different initial conditions of the progenitor galaxies and on
the DM halo. 

The SMBH of Holm 15A is a candidate system for direct imaging of its
sphere of influence. The photon ring radius is
$\sqrt{27}GM_{BH}/c^2 = 2100 \pm \SI{410}{AU}$. At redshift z = 0.055, this corresponds
to an area spanning $18 \pm \SI{3.7}{\mu as}$ on the sky, only slightly smaller
than the current minimum angular resolution of the 
Event Horizon Telescope, $\SI{25}{mas}$ \citep{2019ApJ...875L...4E}. 

\section*{Acknowledgement}
We acknowledge the support by the DFG Cluster of Excellence "Origin and Structure of the Universe".
The dynamical models have been done on the computing facilities of the Computational Center 
for Particle and Astrophysics (C2PAP) and we are grateful for the support by A. Krukau and F. Beaujean through the
C2PAP. We are grateful to Hans B\"ohringer for valuable discussions and suggestions. 
\appendix

\section{Parametric analysis of Wendelstein Photometry}

\subsection{1D-Analysis of the Wendelstein image}
\label{sec:1dphotometry}

The best-fit parameters of the various models we fit to the 1D $i$-band Wendelstein image of Holm 15A in Section \ref{sec:missinglight} are shown in Table \ref{tab:1dcoretab}. In the table, the parameters of the different models are separated into components: Parameters of the Core-S\'ersic
  function $I_{cS}$($r_{b}$, $n$, $\alpha$, $\gamma$, $n_{1}$, $r_{e, 1}$) (see eq. 2 from \citet{2013AJ....146..160R}), outer S\'ersic function $I_{S}$($n_2$, $\mu_{e,2}$ and $r_{e,2}$) (cf. outer S\'ersic components in S\'ersic + S\'ersic models from \citet{Kluge2019}, eq. 11,12) and Nuker function
        $I_N$($r_{b}$, $n$, $\alpha$, $\beta$, $\gamma$) (see eq. 10 from \citet{2007ApJ...662..808L}). 

\begin{table}
\centering
 \begin{tabular}{l|lccccc|l}

      Model & Parameter   &  cSS & cSS($r_{min} = 4 \arcsec$)  &  cSS($r_{min} = 12 \arcsec$)  & N($r_{max} = 20 \arcsec$) & N($r_{max} = 70 \arcsec$) & Units  \\
\hline
 Core-S\'ersic   & $r_{b}$     & $8.96$   & $3.72$   & $5.52$  & $...$ &  $...$  & arcsec      \\
               & $\mu_{b}$   & $20.6$   & $20.1$    & $20.1$  & $...$  & $...$   & mag arcsec$^{-2}$     \\
               & $\alpha$    & $1.71$   & $7.96$   & $2.65$   & $...$ & $...$   &   \\
               & $\gamma$    & $0.09$   & $0.00$  & $0.13$   & $...$   & $...$    &      \\
              & $n_{1}$ &  $5.24$   & $1.38$  & $2.52$  & $...$    & $...$    &          \\
               & $r_{e, 1}$  & $5.20$   & $14.4$    & $14.9$ &     &           \\
\hline
Outer S\'ersic   & $ n_{2}$ & $2.90$   & $5.37$   & $3.30$ &   $...$      &   $...$   & \\
               & $\mu_{e,2}$ & $28.7$   & $28.6$  & $28.8$  &   $...$  &     $...$  & mag arcsec$^{-2}$     \\
               & $r_{e, 2}$  & $917.0$  &$647.3$ & $952.5$  &   $...$  &     $...$   & arcsec          \\
\hline 
Nuker          & $r_{b}$     & $...$    & $...$ & $...$  & $11.09$  &  $8.16$  & arcsec      \\
               & $\mu_{b}$   & $...$    & $...$ & $...$  & $20.9$   & $20.45$  & mag arcsec$^{-2}$     \\
               & $\alpha$    & $...$    & $...$ & $...$   & $1.50$   & $2.30$  & \\
               & $\beta$    & $...$    &$...$  & $...$ & $2.66$   & $2.1$   &  \\
               &  $\gamma$ &  $...$    & $...$  & $...$   & $0.06$   & $0.15$    &     \\
 
\end{tabular}

\caption{Parameters of our best-fit models to the 1D $i$-band Wendelstein image of Holm 15A, separated into components: cSS: Core-S\'ersic + (outer) S\'ersic fit to the light rofile out to \SI{200}{\arcsec} with all parameters fit simultaneously. cSS($r_{min} = 4 \arcsec$): Core-S\'ersic + (outer) S\'ersic fit to the light profile out to \SI{200}{\arcsec} but with the parameters fit in two steps as described in Section \ref{sec:missinglight} with $r_{min} = 4 \arcsec$. cSS($r_{min} = 12 \arcsec$): same as the previous model, but with $r_{min} = 12 \arcsec$.
        N($r_{max} = 20 \arcsec$): Nuker profile fit to the the data within $r_{max} = 20 \arcsec$ with all 5 parameters simultaneously. N($r_{max} = 70 \arcsec$):  same as the previous model, but with $r_{max} = 70 \arcsec$.}
\label{tab:1dcoretab}
\end{table}

\subsection{2D-Analysis of the Wendelstein image}
\label{sec:irafanalysis}
As described in Sec.~\ref{sec:missinglight}, a detailed investigation of the 1D light profile of Holm~15A did not provide strong evidence for a {\it break} radius that separates the inner core from the rest of the galaxy. Here, we describe in detail our 2D fits to the $i$-band image using \textsc{Imfit}
\citep{2015ApJ...799..226E}. Our goal in performing these fits was to better understand
the structure of the unusual core region of Holm~15A. In particular, whether or not a 2D analysis including the ellipticity structure of the galaxy would help in constraining the size of the galaxy's core.

\begin{figure}
 \includegraphics[width=0.9\columnwidth]{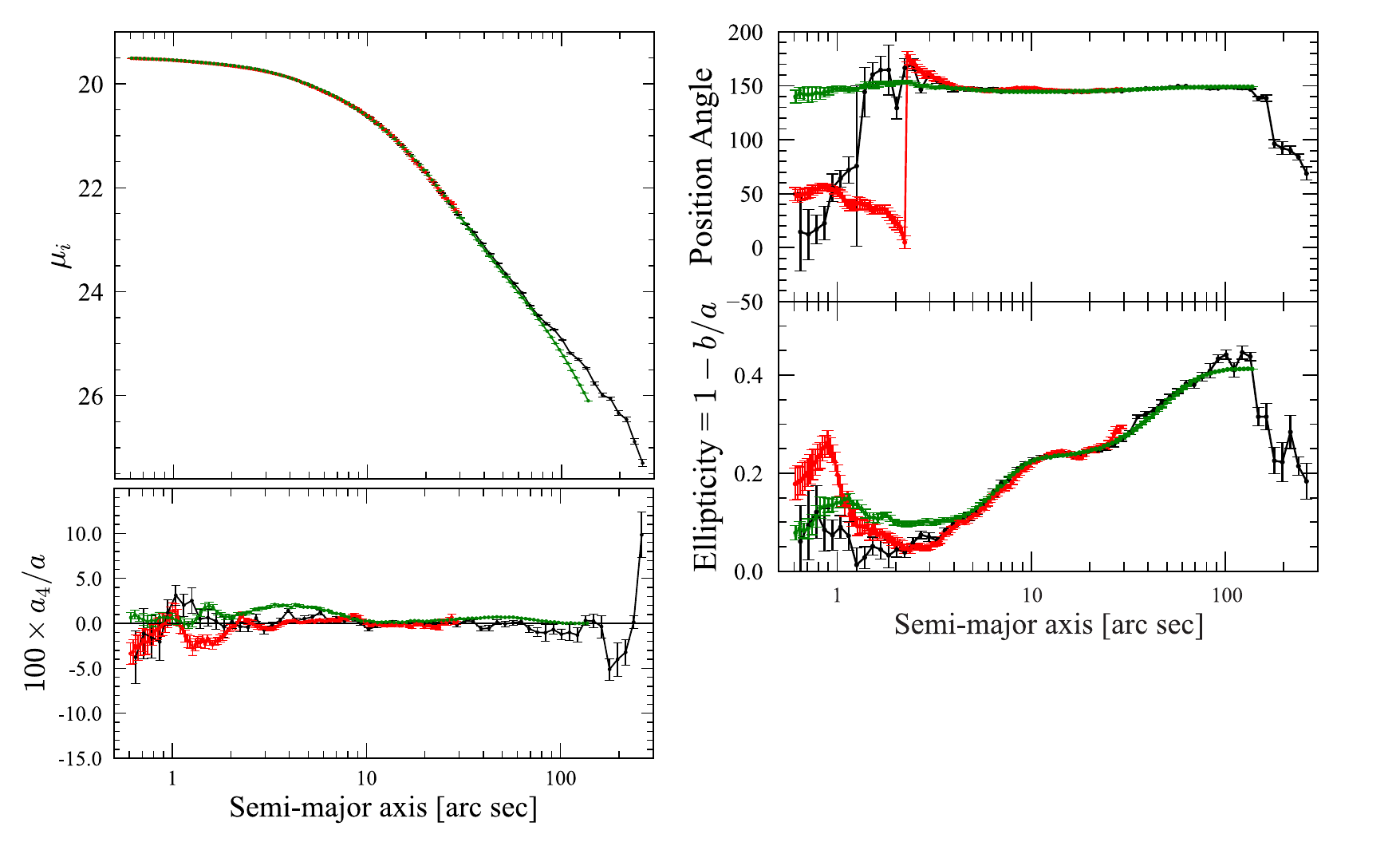}
 \centering
   \caption{Ellipse fits to the isophotes of Holm~15A for our $i$-band
Wendelstein image (black), the red image extracted from our MUSE data cube
(red), and the best-fitting 2D model image (green). From top to bottom, left to right, the panels
show i-band surface brightness, position angle, ellipticity, $a4/a = \sqrt{b/a} * \cos{4\theta}$
parameter versus semi-major axis on a logarithmic scale. The logarithmic scale is shown for the sake of completeness and complementary to the $r^{1/4}$ and linear scale of Figure \ref{fig:holmvlauer}.}
   \label{fig:ellipsfit}
\end{figure}

To have a fully independent analysis, we created ellipse fits to the Wendelstein image using the IRAF task \textit{ellipse} \citep{carter78, jedrzejewski87}, complementary to our analysis 
in Section \ref{sec:photometry}. The surface brightness profile and isophote shape measurements out to 250\arcsec{} (see Figure \ref{fig:ellipsfit}) are fully compatible with the results from the other method (cf. Figures \ref{fig:holmvlauer} and \ref{fig:epsandPA}). 

Beyond about 140\arcsec, the position
angle twists by about 90\degr, and the ellipticity drops from $\sim 0.4$
to $\sim 0.2$ (see Figure \ref{fig:ellipsfit}). Meanwhile, the centers of the fitted ellipses begin
varying by as much as $\sim 15\arcsec$. It is not clear how much of this
represents a real change in the isophotes, e.g. if this is related to a
transition to intra-cluster light, or how much is simply an artifact of the
increasingly low S/N (signal-to-noise ratio). We therefore confine our 2D fitting to $a <
140\arcsec$. In the intervall between  $3\arcsec$ and at least $100\arcsec$, the position angle is remarkably stable, suggesting that Holm~15A might be close to rotational symmetry.
 
Towards the very center, the change in position angle implies that the isophotes start rotating but at the same time the galaxy becomes significantly rounder. 

Because the region of the core is close to circular, the {\it actual} isophotes do not show any
visible twists or distortions.

We find that fitting the image with an inner S\'{e}rsic function that is near-exponential in shape, with a S\'{e}rsic index of $n = 0.99$, and an outer S\'{e}rsic component with $n = 1.48$ results in a
good fit to the Wendelstein data. The inner component is consistent with the S\'{e}rsic + S\'{e}rsic model
listed in \citet{Kluge2019}, though the outer S\'{e}rsic index is smaller. It is also smaller than for our 
core-S\'{e}rsic + S\'{e}rsic models from Section \ref{sec:missinglight} (cf. Table \ref{tab:1dcoretab}). It is however consistent with models from \citet{2011ApJS..195...15D},  who found that Holm~15A's R-band light profile 
is well fit by the sum of two exponential functions (i.e. equiv. to a S\'{e}rsic + S\'{e}rsic model, with both $n \sim 1$). Similar results were obtained from the 2D analysis of a CFHT $r$-band image by \citet{2015ApJ...807..136B}.

However, replacing the inner, exponential-like S\'ersic component with a
core-S\'ersic component, 
did not significantly improve the quality of the fit relative to the core-less model. This reflects the radial trend of the \textit{observed} light profile shown in the bottom panel of Figure \ref{fig:holmvlauer} - The central light profile of Holm~15 is approximately exponential up to $\sim \SI{25}{\arcsec}$.

\begin{figure*}
\centering
 \includegraphics[width=0.5\columnwidth]{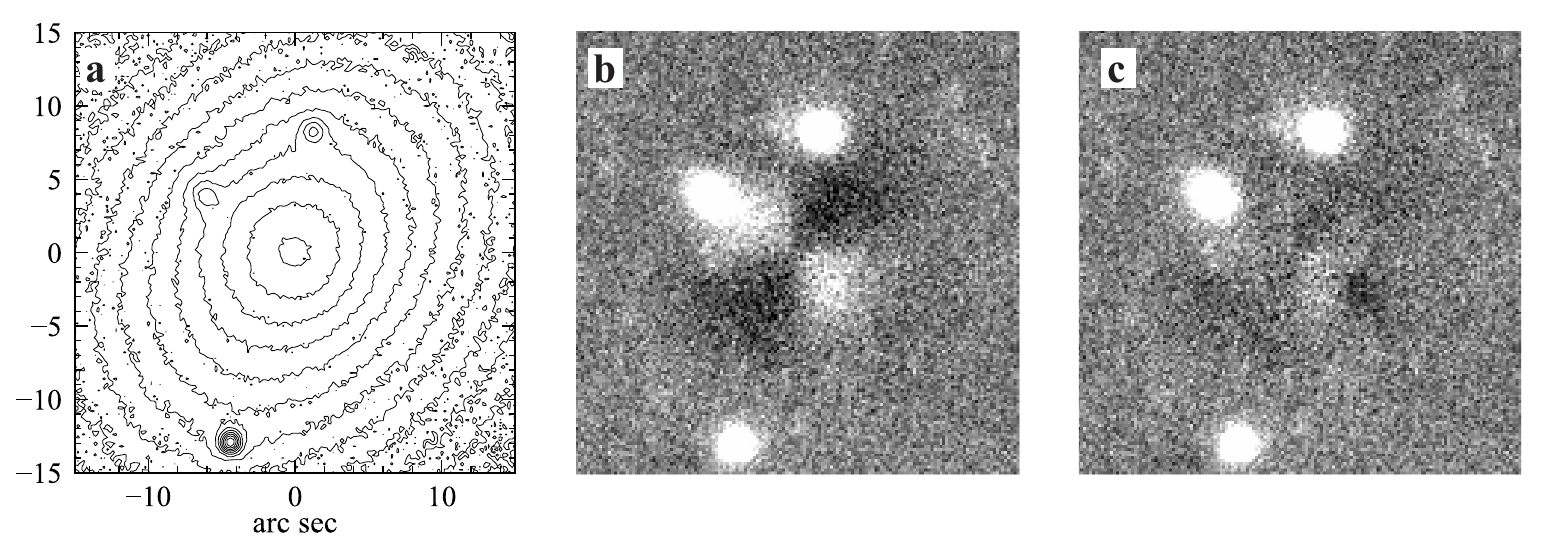}
   \caption{Data and residuals for 2D fits. \textbf{a}: Inner isophotes for
Wendelstein $i$-band image of Holm~15A; peak galaxy intensity
is $\approx 700$ counts/pixel. \textbf{b}: Residuals for the core-S\'ersic + S\'ersic model 
(data $-$ model), plotted on a linear scale from $-25$
(black) to $+25$ (white) counts/pixel. \textbf{c}: Same as for panel~b,
but for the core-S\'ersic + S\'ersic + GaussianRing3D model. 
}
   \label{fig:2dmodelprofiles}
\end{figure*}

Nonetheless, there was still a distinct, bilobed excess in the residual
image from the both the double S\'ersic and the core-S\'ersic + S\'ersic fit, on a scale of $a \sim 4\arcsec$. We therefore
experimented with adding additional components to the model. The best
result was with the GaussianRing3D function of \textsc{Imfit}, which
performs line-of-sight integration for an inclined ring with a Gaussian
radial density and an exponential vertical density. The final result was
a fit with central residuals
which were almost completely lacking in any systematics (see Figure \ref{fig:2dmodelprofiles}). The ``ring'' component has a
semi-major axis of $4\farcs1$, a position angle of 53\degr{} -- almost
perpendicular to the S\'{e}rsic components -- and is intrinsically circular,
viewed at an inclination of 68\degr.  We also note that this may be consistent with
the extra Gaussian-like S\'ersic component (with $n = 0.3$) -- with a
position angle of $\sim 55\degr$ -- which \citet{2015ApJ...807..136B} added to
their 2D fits as a ``corrective'' component. 
We emphasize that this is a purely
empirically chosen function which produces approximately the right excess light
to minimize the residuals; it is not necessarily evidence for an actual
inclined ring. The parameters of the best-fit 2D model are listed in
Table \ref{tab:phototab}.

In summary, while the 2D analysis provides somewhat more stable fit parameters, it confirms the results from Sec.~\ref{sec:missinglight}, in particular the lack of a clear break radius.
In the 2D analysis, we assume a spatially constant
flattening for each individual component. This might imply that the components simply trace the structure of the ellipticity profile of the galaxy and this, in turn,
could explain why the parameters of the 2D fits turn out more stable than in the 1D analysis. It is not clear at the moment
how much or which physical information is encoded in the ellipticity profile of the galaxies. Likewise, it is not clear how physically significant the extra-light ring might be, which has a total luminosity comparable to the expected amount of stars ejected from the center by a SMBH binary, i.e. the extra light is of a similar order $\sim 0.5 \times 10^{10} L_{\odot}$ as the missing light determined in Section \ref{sec:missinglight}. 
Comparing the distribution of stars in Holm~15A to those of other cored ETGs (cf. Figure \ref{fig:holmvlauer}) makes clear
that Holm~15A is not only characterised by an extreme \textit{deficit} of light in the inner core but also by an 
\textit{excess} of light adjacent to the core. This light "excess", however, extends well beyond the extra-light ring (roughly out to $20 \arcsec$).

\begin{table}
\centering
 \begin{tabular}{c|c|c|c}
Component       & Parameter & Value               & units \\
\hline
Core-S\'ersic     & PA         & $141.9 \pm 0.2$    & deg \\
                & $\epsilon$ & $0.187 \pm 0.002$  &     \\
                & $n$        & $0.965 \pm 0.005$  &     \\
                & $I_{b}$    & $20.040 \pm 0.012$    & mag~arcsec$^{-2}$ \\
                & $r_{e}$    & $12.87 \pm 0.04$   & arc sec \\
                & $r_{b}$    &  $2.57 \pm 0.05$   & arc sec \\
                & $\alpha$   & $12.15 \pm 4.1$    &  \\
                & $\gamma$   & $0.096 \pm 0.007$  &  \\
S\'ersic          & PA         & $149.0 \pm 0.1$    & deg \\
                & $\epsilon$ & $0.413 \pm 0.003$  &     \\
                & $n$        & $1.69 \pm 0.03$    &     \\
                & $I_{e}$    & $24.035 \pm 0.016$   & mag~arcsec$^{-2}$ \\
                & $r_{e}$    & $60.67 \pm 0.48$   & arc sec \\
GaussianRing3D  & PA         & $52.1 \pm 0.9$     & deg \\
                & inclination & $81.8 \pm 1.5$    & deg \\
                & $J_{0}$    & $1.08 \pm 0.03$    & counts~pixel$^{-3}$ \\
                & $a$        & $4.37 \pm 0.07$     &  arc sec \\
                & $\sigma$   & $1.76 \pm 0.05$     & arc sec \\
                & $h_{z} $   & $2.78 \pm 0.10$     & arc sec \\
   
\end{tabular}

\caption{Best-fit \textsc{Imfit} model for the $i$-band image of Holm
15A. Column 1: component used in fit. Column 2: parameter. Column 3:
best-fit value for parameter and $1-\sigma$ confidence limits from 200
rounds of bootstrap resampling. Column 4: units. Note that for the
GaussianRing3D component, we fixed the ring PA and ellipticity to both
be zero, so these are not listed in the table.}
\label{tab:phototab}
\end{table}

\section{Stellar kinematics}
\subsection{Kinematics of Holm~15A compared to MASSIVE survey ETGs}
\label{sec:kincompare}
To better understand Holm~15A's place among other known massive ETGs we will compare it's stellar kinematics
to ETGs from the MASSIVE survey \citep[][and subsequent MASSIVE survey papers]{2014ApJ...795..158M}.
\\Characterizing Holm15~A's velocity dispersion profile, $\sigma(r)$ (see Section \ref{sec:spectra})
by fitting a combined power-law profile as 
suggested by \citet{2018MNRAS.473.5446V} in their study of the 90 ETGs of the MASSIVE survey, we find an 
inner logarithmic slope $\gamma_{inner} = -0.017 \pm 0.007$ of the $\sigma$ profile at $\sim \SI{2}{kpc}$ and an
outer logarithmic slope $\gamma_{outer} = 0.029 \pm 0.009$ at $\sim \SI{20}{kpc}$. Roughly $90 \%$ of BCGs in 
the MASSIVE survey have $\gamma_{inner} \leq 0$ and $\sim 60 \%$ with $\gamma_{outer} \geq 0$.
Moreover, for the eleven 
most massive BCGs in their sample with $M_{*} \sim 10^{12} \si{M_{\odot}}$, $\gamma_{inner} \leq 0$ and $\gamma_{outer} \geq 0$
for all except one. The scatter in $\gamma_{inner}$ and $\gamma_{outer}$ between these eleven most massive BCGs is quite high, 
$\overline{\gamma}_{inner} = -0.040 \pm 0.055$ and $\overline{\gamma}_{outer} = 0.088 \pm 0.084$.
Nonetheless, statistically, their overall rather flat $\sigma(r)$ profiles are similar to the one in Holm~15A,
even though the galaxy's average 
velocity dispersion within one effective radius $\sigma_{e} \sim \SI{340}{km/s}$ is slightly higher compared to 
these 
BCGs $ \sim \SI{300} {km/s}$.
\\The parameter $h_4$, in our measured kinematic profile starts out at $\sim 0.07$ within $\SI{2}{kpc}$ and rises to $\gtrsim 0.1$ along 
the major axis towards the edges of the MUSE FOV. All 11 of the most massive MASSIVE BCGs share this trend of $h_{4}> 0$ over 
their respective radial coverage and all but one have positive $h_{4}$ gradients towards larger radii. Similarly as with 
$\sigma$, average values for $h_{4}$ within $r_{e}$ are larger for Holm~15A, $h_{4, e} \sim 0.08$ than for those other BCGs 
where  $h_{4, e} \lesssim 0.06$. 
Essentially all galaxies in the MASSIVE sample with $h_{4, e} > 0.05$ (BCG or not) have 
within the central \SI{2}{kpc}
super-solar $[\alpha/Fe] > 0.2$ and most galaxies with $h_{4, e} > 0$ have $[Fe/H] \leq 0$ \citep{2019arXiv190101271G}. 
\\Using stellar population models of Lick indices \citep{2003MNRAS.339..897T, Maraston+11} we find abundance ratios in good agreement with these in Holm~15A:
$[\alpha/Fe] = 0.25 \pm 0.03$ and $[Fe/H] = -0.011 \pm 0.008$.
\\Overall, we find stellar kinematics in Holm~15A similar to those of other known massive ETGs. Indeed,
from a stellar-kinematic point of view we find no indication that Holm~15A is anything other than a 
higher-mass extrapolation of known
massive ETGs in the local universe, the vast majority of which is cored \citep[e.g.][]{2007ApJ...664..226L,  2013MNRAS.433.2812K,
2013ARA&A..51..511K}.
\subsection{Non-parametric kinematics compared to Gauss-Hermite polynomials}
\label{sec:nonParavPara}
\begin{figure}
      \includegraphics[width=0.7\columnwidth]{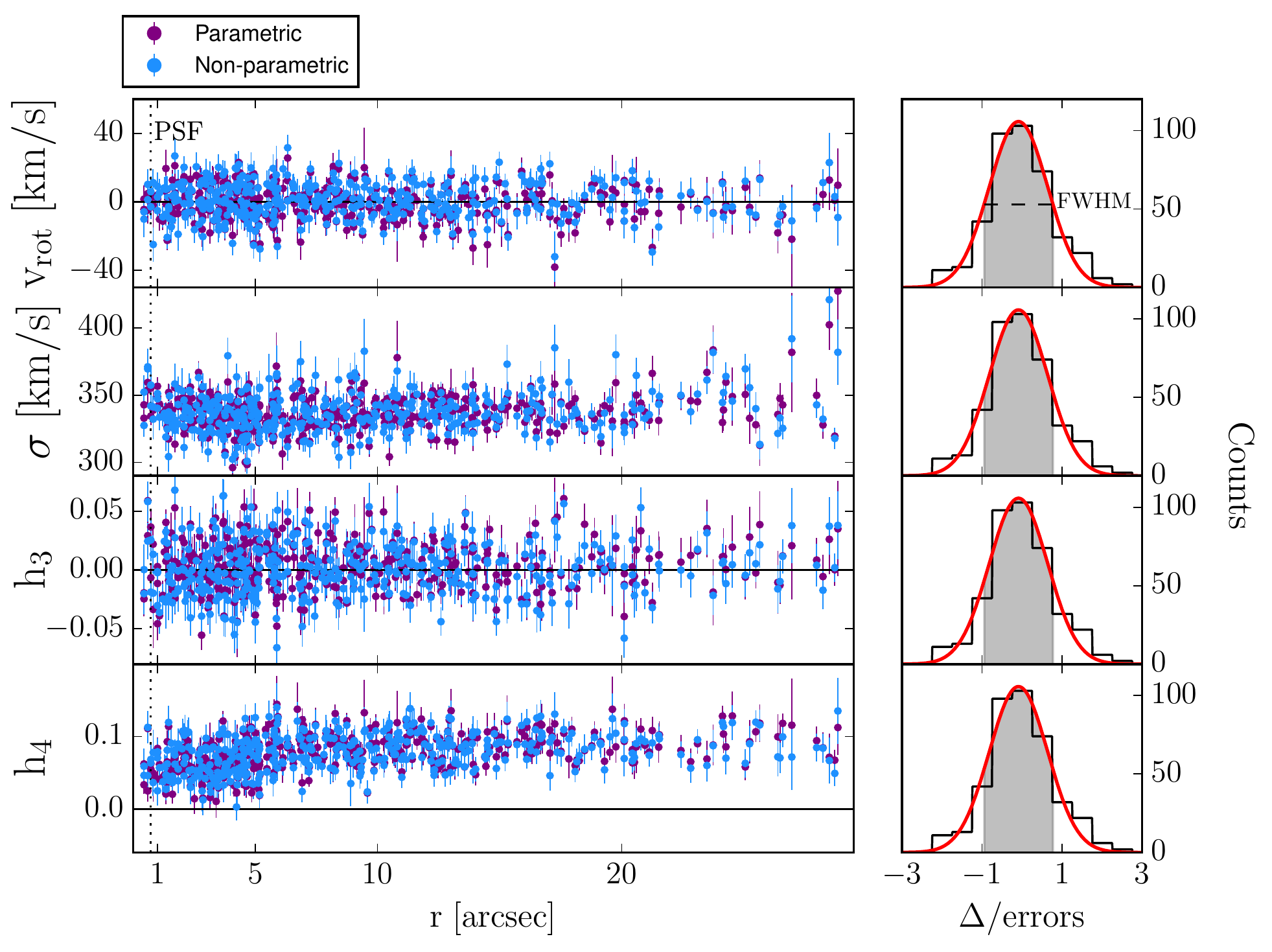}
	\centering
       \caption{Left column: Resulting kinematic profiles over radius of the two kinematic measurements
        performed in this study, one using pPXF (purple points) and one with our own non-parametric code (blue points). 
        Panels show, from top to bottom, radial profiles for $v_{rot}$, $\sigma$, $h_3$ and $h_{4}$, including statistical
        uncertainties.
        For this plot non-parametric LOSVDs
        were fitted with a Gaussian times third to forth order Gauss-Hermite polynomials.
        In our final modeling non-parametric LOSVDs are used, but 
        these parameters still allow us to showcase the kinematic structure of Holm~15A.
        Right column, from top to bottom: Corresponding distributions of the difference $\Delta$ (black) between 
        pPXF and non-parametric LOSVD Gaus-Hermite parameters over the statistical uncertainties of the pPXF values. 
         Each distribution is fit with a Gaussian (red) with the FWHM of each distribution indicated by gray shaded areas.}
	\label{fig:wingprofile}
\end{figure}

We compare the non-parametric stellar
kinematics we measured with our own code with those derived parametrically with pPXF.
This is illustrated in Figure \ref{fig:wingprofile} for all
bins of our FOV (i.e. LOSVDs from all quadrants). Both kinematic profiles are, for the purpose of illustration, 
parameterized via Gaussian times third to forth order Gauss-Hermite polynomials. As the distribution of 
differences in the right column of the figure show, both methods agree within their uncertainties.

\section{Non-parametric dynamical modeling: escape velocities}
\label{sec:escdyn}

\begin{figure}
      \includegraphics[width=0.5\columnwidth]{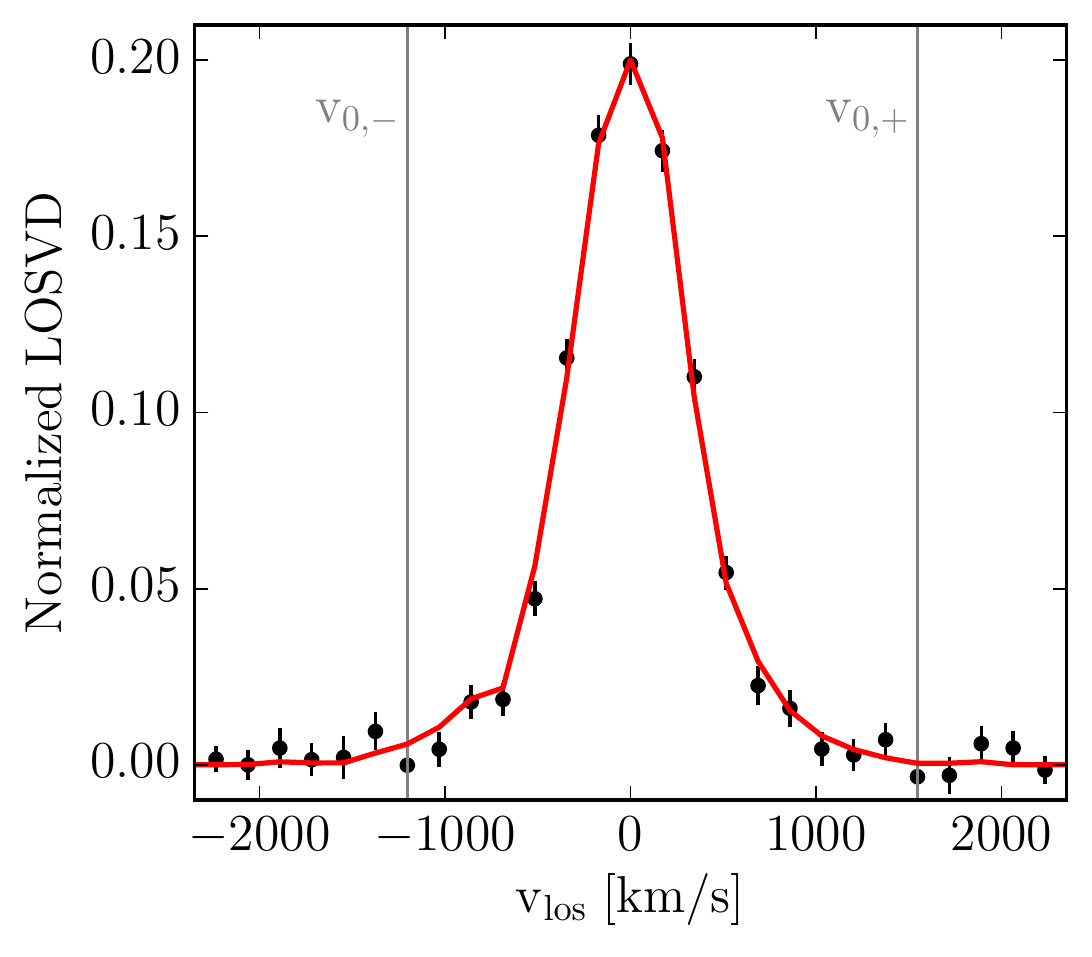}
	\centering
        \caption{Example of a non-parametric fit of our dynamical model (red line) to a
	  non-parametric LOSVD from the center of Holm~15A (black points). The cutoff velocities
	  of the LOSVD are marked as gray, vertical lines. 
           }
	\label{fig:nonparalfit}
\end{figure}

Here, we will briefly discuss the connection between the wings of the
observed line-of-sight velocity distributions on the one side and the
mass distribution and orbital structure on the
other. Figure~\ref{fig:nonparalfit} shows an example of a non-parametric
LOSVD measured near the center of Holm~15A together with the
corresponding LOSVD from our best-fit dynamical model. We define the
cutoff velocity $v_0$ of any LOSVD as the mean $v_0 = (v_{0,+} +
v_{0,-})/2$. If $v_\mathrm{peak}$ denotes the line-of-sight velocity
at which the LOSVD peaks, then $v_{0,+}$ is the smallest zero of the
LOSVD for $v_\mathrm{los} > v_\mathrm{peak}$ and $v_{0,-}$ is the absolute value of the 
largest zero of the LOSVD for $v_\mathrm{los} < v_\mathrm{peak}$,
respectively. For Holm~15A this definition is sufficient since there is
almost no detectable rotation and the LOSVDs are largely symmetric with respect to
$v_{peak}$.
For the LOSVD in Figure~\ref{fig:nonparalfit} we measure $v_0 \sim
\SI{1375}{km/s}$.

\begin{figure}
      \includegraphics[width=0.7\columnwidth]{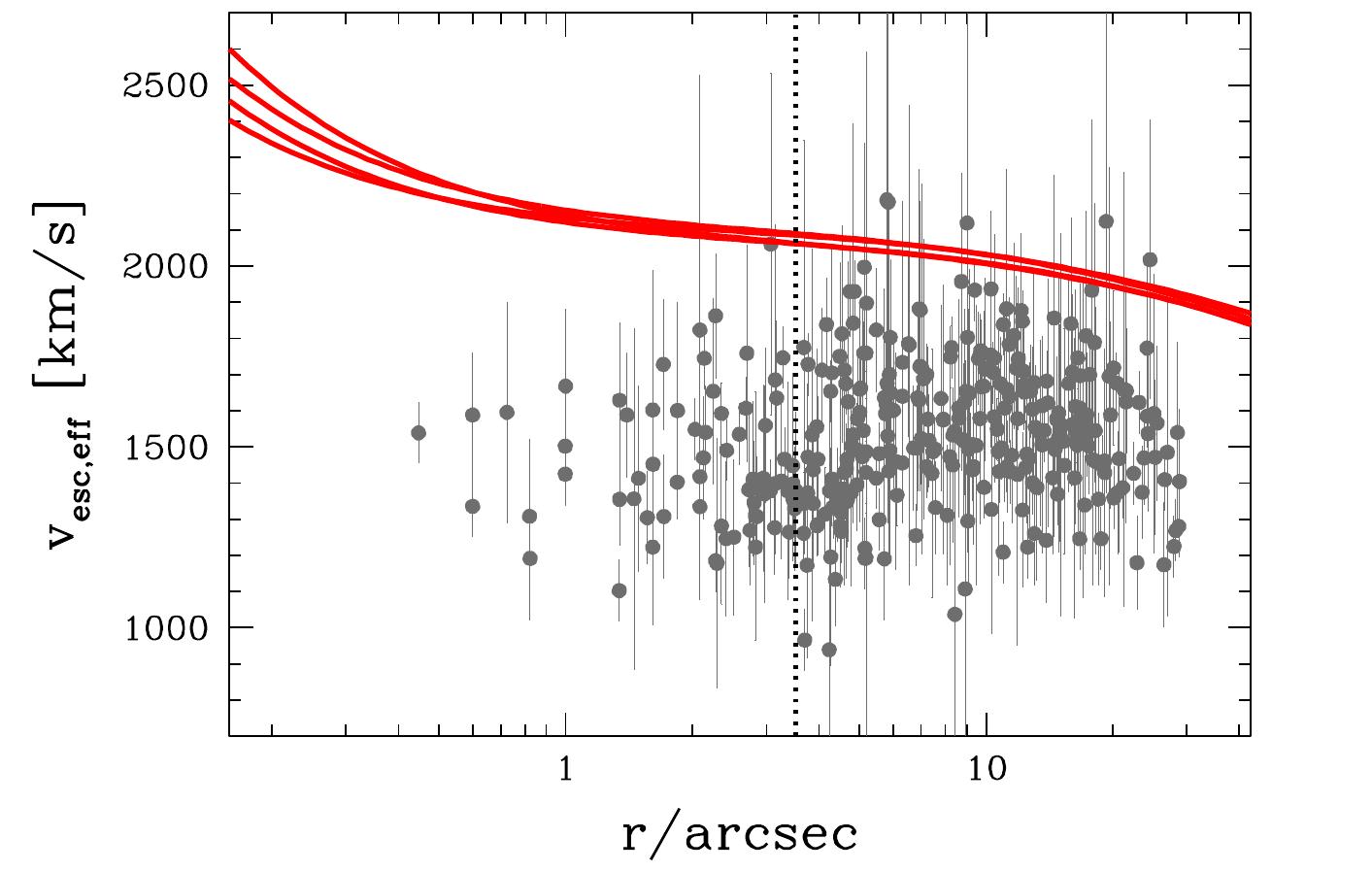}
	\centering
        \caption{Effective escape velocities measured for every LOSVD of our FOV (grey points)
	  and the escape velocities of the gravitational potential of our best fit  dynamical model 
	  of Holm~15A
	  (4 red lines, one for each quadrant) versus radius. The vertical dotted line indicates the SOI of the black hole.
           }
	\label{fig:vescape}
\end{figure}

Figure \ref{fig:vescape} shows all the measured cutoff velocities
$v_0$ from our MUSE observations together with the escape velocity
curves $v_\mathrm{esc}(r)$ of the four best-fit models for the four
quadrants of the galaxy. Here, we define $v_\mathrm{esc}$ relative to
the maximum radius that is sampled by the orbit library. The uncertainties of the cutoff velocities
are measured via the difference between values of $v_{0}$ determined from $LOSVD(v_{los}) + \Delta LOSVD(v_{los})$ and
$LOSVD(v_{los}) - \Delta LOSVD(v_{los})$.
Outside the
core ($r \ga 5 \, \mathrm{kpc}$), the best-fit $v_\mathrm{esc}(r)$
curves follow closely the maximum observed cutoff velocities
$v_0$. This is expected in a radially anisotropic system where a
significant number of stars is populated on weakly bound, radially
extended orbits. The less bound and the more radial the orbit is, the
closer the orbital velocity gets to $v_\mathrm{esc}$. Indeed, outside
the core region, our best-fit models become increasingly radially
anisotropic (cf. Figure \ref{fig:ani}).

The situation changes towards the center of the galaxy, where the
gravitational well is deepest. The observed cutoff velocities {\it
decrease} at small radii, whereas the escape velocity necessarily
increases. This can only be explained as an anisotropy effect: inside
the sphere-of-influence of the central black hole (indicated by the vertical
line), the orbit distribution becomes tangential (cf. figure
\ref{fig:ani}). Since only stars on the most radial orbits can move
with velocities up to the escape velocity and those stars are missing,
the LOSVDs do not extend to $v_\mathrm{esc}$ anymore but vanish at
smaller velocities.

The uncertainties in the observed cutoff velocities are large (due to
the noise in the wings of the LOSVDs). This is indicated by the large
scatter in values of $v_{0}$.  However, the figure clearly
demonstrates the importance of the information contained in the wings
of the LOSVDs for both the gravitational potential as well as for the
orbital structure.

\bibliography{bibliography} 
\end{document}